\documentclass[twocolumn,twocolappendix]{aastex631}

\usepackage{physics}
\usepackage{amsmath}
\usepackage{tensor}
\usepackage{tabu}
\usepackage{booktabs}
\usepackage{subfigure}
\usepackage{hyperref}
\usepackage{ulem}
\usepackage{enumerate}
\usepackage{multirow}
\usepackage{CJK}

\newcommand{\sect}{Section~}
\newcommand{\sects}{Sections~}
\newcommand{\figu}{Figure~}

\newcommand{\tab}{Table~}

\newcommand{\eq}{Equation~}

\newcommand{\athenapp}{\texttt{Athena++}}
\newcommand{\athenak}{\texttt{AthenaK}}

\shorttitle{Cyclic Zoom}
\shortauthors{Guo et al.}
\graphicspath{{./}{figures/}}

\begin{document}

\title{Cyclic Zoom: Multiscale GRMHD Modeling of Black Hole Accretion and Feedback}

\begin{CJK*}{UTF8}{gbsn}

\correspondingauthor{Minghao Guo}
\email{mhguo@princeton.edu}

\author[0000-0002-3680-5420]{Minghao Guo (郭明浩)}
\affiliation{Department of Astrophysical Sciences, Princeton University, Princeton, NJ 08544, USA}

\author[0000-0001-5603-1832]{James M. Stone}
\affiliation{Department of Astrophysical Sciences, Princeton University, Princeton, NJ 08544, USA}
\affiliation{School of Natural Sciences, Institute for Advanced Study, 1 Einstein Drive, Princeton, NJ 08540, USA}
\author[0000-0001-9185-5044]{Eliot Quataert}
\affiliation{Department of Astrophysical Sciences, Princeton University, Princeton, NJ 08544, USA}
\author[0000-0001-5976-4599]{Volker Springel}
\affiliation{Max-Planck-Institut f{\"u}r Astrophysik, Karl-Schwarzschild-Stra{\ss}e 1, D-85740 Garching bei M{\"u}nchen, Germany}

\begin{abstract}
We present a ``cyclic zoom'' method to capture the dynamics of accretion flows onto black holes across a vast range of spatial and temporal scales in general relativistic magnetohydrodynamic (GRMHD) simulations. In this method, we cyclically zoom out (derefine) and zoom in (refine) the simulation domain while using a central mask region containing a careful treatment of the coarsened fluid variables to preserve the small-scale physics, in particular the magnetic field dynamics. The method can accelerate GRMHD simulations by $\gtrsim 10^5$ times for problems with large-scale separation. We demonstrate the validity of the technique using a series of tests, including spherically symmetric Bondi accretion, the Blandford-Znajek monopole, magnetized turbulent Bondi accretion, accretion of a magnetized rotating torus, and the long-term evolution of an accreting torus about both Schwarzschild and Kerr black holes. As applications, we simulate Bondi and rotating torus accretion onto black holes from galactic scales, covering an extremely large dynamic range. In Bondi accretion, the accretion rate is suppressed relative to the Bondi rate by $\sim(10r_\mathrm{g}/r_\mathrm{B})^{1/2}$ with a feedback power of $\sim 0.01 \dot{M} c^2$ for vanishing spin, and $\sim 0.1 \dot{M} c^2$ for spin $a\approx0.9$. In the long-term evolution of a rotating torus, the accretion rate decreases with time as $\dot{M}\propto t^{-2}$ on timescales much longer than the viscous timescale, demonstrating that our method can capture not only quasi-steady problems but also secular evolution. Our new method likewise holds significant promise for applications to many other problems that need to cover vast spatial and temporal scales.
\end{abstract}


\section{Introduction} \label{sec:intro}

Supermassive black holes (SMBHs), harbored in the nuclei of almost all galaxies, correlate with properties of their hosts~\citep{Magorrian1998AJ....115.2285M, Ferrarese2000ApJ...539L...9F, Gebhardt2000ApJ...539L..13G, Kormendy&Ho2013ARA&A..51..511K}. How these black holes accrete gas from the ambient background and feed mass, momentum, and energy back into their environments remain crucial unsolved problems. Accurately modeling feeding and feedback from galactic to event horizon scales is a formidable task; the involved spatial scales span nearly nine orders of magnitude (from mpc to Mpc) and need to be resolved over an extended time period~\citep{Gaspari2020NatAs...4...10G}. Due to limits of computational power, special prescriptions (``sub-grid'' models) to model accretion flows near the SMBHs and their feedback are widely adopted in galactic and cosmological simulations~\citep[e.g.,][]{Li&Bryan2014ApJ...789..153L, Weinberger2017MNRAS.465.3291W, Weinberger2025arXiv250213241W, Tremmel2017MNRAS.470.1121T, Pillepich2018MNRAS.473.4077P, Springel2018MNRAS.475..676S, Ni2022MNRAS.513..670N, Ni2024arXiv240910666N, Su2023MNRAS.520.4258S, Koudmani2024MNRAS.532...60K}.

On the other hand, general relativistic magnetohydrodynamic simulations (GRMHD) that resolve event horizon scales~\citep{Gammie2003ApJ...589..444G, Narayan2012MNRAS.426.3241N, Narayan2022MNRAS.511.3795N, Porth2019ApJS..243...26P, White2020ApJ...891...63W, Yang2021ApJ...914..131Y, Chael2025MNRAS.537.2496C} often adopt idealized initial conditions, e.g., a torus in hydrodynamic equilibrium~\citep{Fishbone1976ApJ...207..962F, Kozlowski1978A&A....63..209K}. Modeling infall and accretion on galactic scales from the Bondi radius~\citep{Bondi1952MNRAS.112..195B} or even farther out may help to construct a more realistic model of black hole accretion at smaller radii~\citep{Yuan2014ARA&A..52..529Y, Gaspari2020NatAs...4...10G}. Connecting the large and small scales is also critical for developing more physical models of black hole growth and feedback in cosmological simulations that lack the physics or resolution to follow the gas at small radii~\citep{Hopkins&Quataert2010MNRAS.407.1529H, Hopkins&Quataert2011MNRAS.415.1027H, Li&Bryan2014ApJ...789..153L}.

Recent years have seen considerable efforts to link these scales in various environments using different techniques, including direct simulations assuming smaller scale separation~\citep[e.g.,][]{Lalakos2022ApJ...936L...5L, Kaaz2023ApJ...950...31K, Olivares2023A&A...678A.141O, Lalakos2024ApJ...964...79L, Galishnikova2025ApJ...978..148G}, ``zoom-in'' using nested meshes~\citep[e.g.,][]{Ressler2018MNRAS.478.3544R, Ressler2020ApJ...896L...6R, Guo2023ApJ...946...26G, Guo2024ApJ...973..141G}, ``super-Lagrangian'' or ``hyper-refinement'' methods~\citep[e.g.,][]{Angles2021ApJ...917...53A, Hopkins2024OJAp....7E..18H, Hopkins2024OJAp....7E..19H, Hopkins2025OJAp....8E..48H}, or remapping between different simulations~\citep{Kaaz2025ApJ...979..248K}. 

Simply using GRMHD simulations that resolve horizon scales to compute the feedback from the event horizon to galactic scales is generally infeasible due to the very restrictive time step constraints, which limit the evolutionary time that can be modeled.  Recently, \citet{Cho2023ApJ...959L..22C, Cho2024ApJ...977..200C} have introduced a ``multi-zone'' method that greatly accelerates such calculations while passing information both from large to small scales and vice versa. Inspired by this work, we have developed a very similar but nevertheless new technique we term the ``cyclic zoom'' method. In our approach, we cyclically zoom in and zoom out the simulation domain using a carefully designed mask in the inner regions that preserves information about the fluid quantities on small scales. Moreover, the method evolves the magnetic field within the mask to preserve the divergence-free constraint using inductive electric fields that also preserve information from small scales.

The rest of this article is organized as follows. In \sect\ref{sec:method} we describe the numerical method and in particular how we formulate the mask region. In \sect\ref{sec:validation} we illustrate the validity of the method using a series of tests. In \sect\ref{sec:results} we then presents results for Bondi and torus accretion over scales that are simply infeasible to model using traditional approaches on current computational resources. In \sect\ref{sec:discussion}, we discuss the implications of our results. Finally, we conclude in \sect\ref{sec:summary}.

Throughout the paper, we define the gravitational radius as $r_\mathrm{g}=GM/c^2$ and the gravitational time as $t_\mathrm{g}=r_\mathrm{g}/c$ assuming a unit system where $GM=c=1$. As is customary, Greek indices run through $[t,x,y,z]$ and Roman indices span $[x,y,z]$.

\section{Methodology} \label{sec:method}
We perform GRMHD simulations using \athenak{} \citep{Stone2024arXiv240916053S}, a performance portable version of the \athenapp~\citep{Stone2020ApJS..249....4S} code. \athenak{} supports a variety of reconstruction methods, Riemann solvers, and integrators for solving the GRMHD equations in a Cartesian Kerr-Schild grid. In our simulations, we adopt the piecewise parabolic spatial reconstruction method (PPM4), the HLLE Riemann solver, and the RK2 time integrator to solve the GRMHD equations. The adaptive mesh refinement (AMR) in \athenak{} allows us to flexibly achieve a high resolution and good performance over an extremely large dynamic range. We apply the first-order flux correction algorithm \citep[described by][]{Lemaster2009ApJ...691.1092L} in the rare cases that the higher-order fluxes would lead to negative density or pressure. The code framework, equations we solve, and the algorithms implemented in \athenak{} are described in detail by a series of previous papers including \citet{White2016ApJS..225...22W}, \citet{Stone2020ApJS..249....4S}, and \citet{Stone2024arXiv240916053S}.

In the GRMHD module, the conserved variables are $U\in\{\rho u^t, \rho u^t+\tensor{T}{^t_t}, \tensor{T}{^t_i}\}$, i.e, coordinate-frame mass density with rest-mass density $\rho$ and the stress-energy tensor, which has components
\begin{equation}
    \tensor{T}{^\mu^\nu}=\left(\rho+u+p+b^2\right)u^{\mu}u^{\nu}+\left(p+\frac{b^2}{2}\right)\tensor{g}{^\mu^\nu}-b{^\mu}b{^\nu},
\end{equation}
where $u$ is the fluid-frame internal energy density, $p=(\gamma_\mathrm{ad}-1)u$ is the fluid-frame gas pressure with $\gamma_\mathrm{ad}$ the adiabatic index, $u^\mu$ is the coordinate-frame four-velocity, $b^2=b^\mu b_\mu$ with $b^\mu$ the fluid-frame magnetic field four-vector, and $\tensor{g}{^\mu^\nu}$ is the metric. The primitive variables are $W\in\{\rho, u, \tilde{u}^i\}$ where $\tilde{u}^i$ are normal-frame spatial velocity components, related to the coordinate-frame velocity via $u^i=\tilde{u}^i-\alpha\gamma \tensor{g}{^t^i}$, where $\alpha=(-g^{tt})^{-1/2}$ is the lapse and $\gamma$ is the normal-frame Lorentz factor. The coordinate-frame magnetic field $B^i$, defined as the time component of the dual of the electromagnetic field tensor $^*F^{\mu\nu}$, is subject to the divergence-free (no-monopole) constraint
\begin{equation}
    \frac{1}{\sqrt{-g}}\partial_i(\sqrt{-g}B^i)=0,
\end{equation}
where $g=\det(g_{\mu\nu})$. This reduces to $\div\boldsymbol{B}=0$ if $\sqrt{-g}=1$, as in the Cartesian Kerr-Schild coordinates we adopt here.

\begin{figure*}[ht!]
\includegraphics[width=0.49\linewidth]{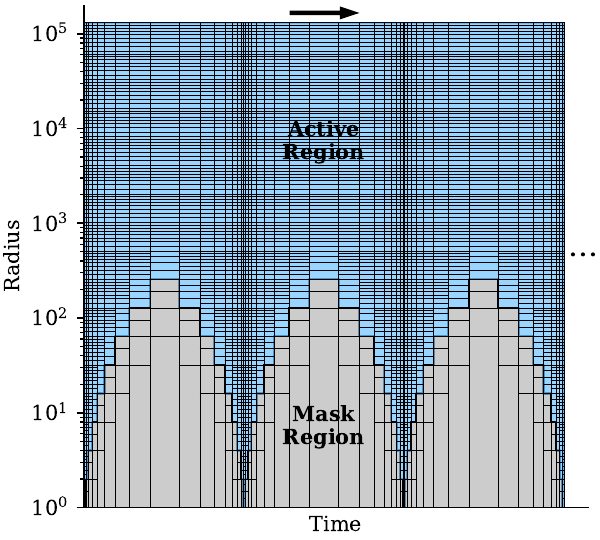}
\includegraphics[width=0.49\linewidth]{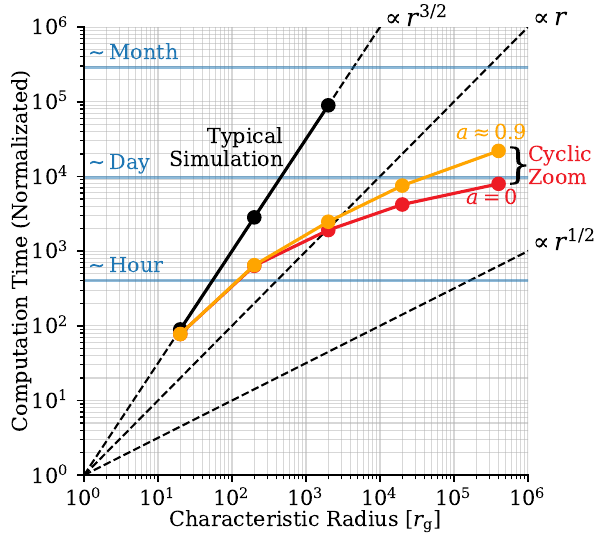}
\includegraphics[width=\linewidth]{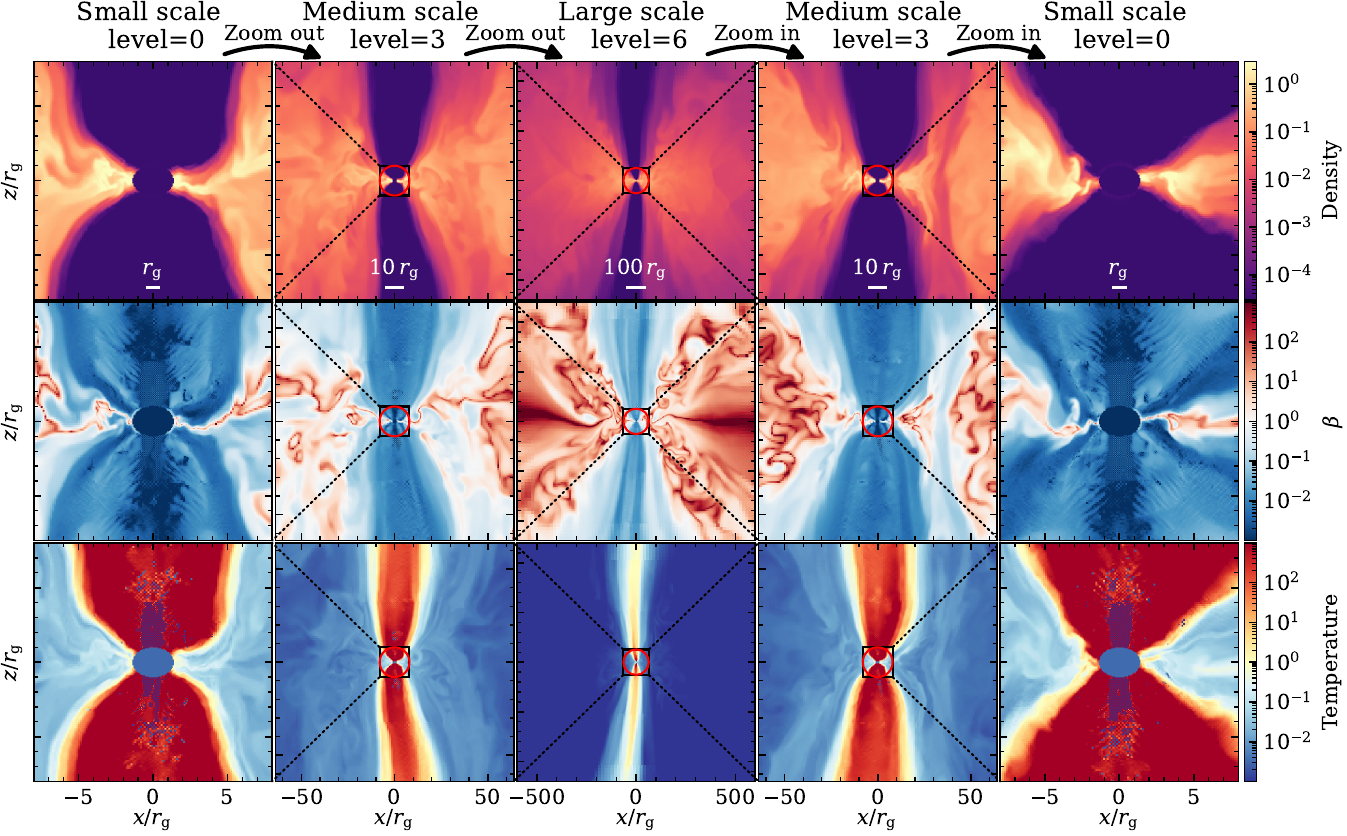}
\caption{Top left: schematics of the cyclic zoom method in a spacetime diagram. Blue marks the active region, and the gray marks mask region. Runtime is shown along the $x$-axis (not to scale). The thin lines illustrate the relative spatial resolution and temporal resolution. Here, only three ``$\Lambda$-cycles'' are shown, but a complete simulation typically consists of tens to hundreds of $\Lambda$-cycles. Top right: Normalized computation time as a function of characteristic radius (Bondi radius) over one characteristic timescale for Bondi accretion in GRMHD simulations (black dots) and cyclic zoom runs with black hole spin $a=0$ (red) and $a=0.9375$ (orange). Blue lines mark typical timescales of an hour, a day, and a month in real time for one GPU node. Typical simulations follow a scaling of $\propto r^{3/2}$, which quickly becomes prohibitive for growing characteristic radius while the cyclic zoom run is relatively insensitive to the characteristic radius. Bottom: illustration of one $\Lambda$-cycle in the cyclic zoom method for accretion of an \citet{Fishbone1976ApJ...207..962F} torus in magnetically arrested state onto a black hole with spin $a=0.9375$ when the system is quasi-steady. We only show levels $i=0,3,6,3,0$ for clarity. The red circles mark the boundary between the mask region in the center and the active region outside. The mask region can maintain the small-scale hydrodynamic and magnetic information and drive the feedback from the central region. It also recovers the accretion flow around the black hole when zooming back into the smallest scale.
\label{fig:method}}
\end{figure*}

\subsection{Cyclic Zoom Method}
The challenge of simulating accretion and feedback from SMBHs to galactic scales is the vast range of spatial and temporal scales. To tackle this problem, we repeatedly zoom out (derefine) and zoom in (refine) the simulation domain, which we denote as the ``cyclic zoom'' method. The spatial resolution is correspondingly decreased (increased) when we zoom out (in) using AMR to alleviate the time step constraints while keeping the same relative resolution $\Delta x/x$ for the region of interest. A mask region in the center is used to preserve the small-scale physics. 
\figu\ref{fig:method} (top left) illustrates the cyclic zoom method in a spacetime diagram with relative spatial and temporal resolution. It behaves like a ``$\Lambda$-cycle'' in the spacetime diagram (or ``V-cycle'', depending on what we choose as the starting point of the simulation). A complete simulation typically consists of tens to hundreds of $\Lambda$-cycles.

The key to the method is the mask region, when we do not resolve the event horizon, as well as how we pass information across the boundaries between the mask region and the active region. It is crucial because, without a mask region or any special treatment, the system cannot be aware of small-scale information such as accretion or feedback. In the Cartesian coordinates we use, the mask region (the gray region in \figu\ref{fig:method}) is a region of cells within a certain radius $r_{\mathrm{z},i}\equiv 2^i\,r_\mathrm{g}$ for level $i$, where $i\in\{0,1,...,(n-1)\}$. In the current resolution we use, we resolve the mask region by $r_{\mathrm{z},i}=16\Delta x_{i}$, which translates to $\gtrsim 10^4$ cells in the mask region. As we zoom out, it is effectively the boundary of the mask region that prescribes an inner boundary condition for the rest of the domain. At each level, making the mask region larger at fixed $\Delta x$ resolves the boundary better and in this sense can feed small-scale information more faithfully into the coarser simulation. On the other hand, keeping a fixed mask region is an approximation and not accurate in detail, so one may need to avoid using a large-volume mask region. In addition, if the mask region is too large, the time step will be limited again due to the smaller cell size relative to the active region. So, in practice, one should aim for a compromise. After trial and error, we find that a resolution of $r_{\mathrm{z},i}/\Delta x_{i}\gtrsim 10$ for the mask region is satisfactory. Due to the constraint of computational resources, we have not yet tested a larger size of mask region, which we defer to future work when presenting higher-resolution simulations.

The simulation is simply a standard GRMHD simulation without any additional treatment when $i=0$. When $i>0$, to maintain small-scale information, we fix the hydrodynamic variables in the central mask region within $r<r_{\mathrm{z},i}$ at level $i$ to be the coarse-grained hydrodynamic variables from the finer grid at level $i-1$. 
That is, every time step, we reset the primitive hydrodynamic variables $W$ to be a constant $W_0$, which is converted from the conservative hydrodynamic variables $U_\mathrm{hydro}\in\{\rho u^t, \rho u^t + \tensor{T}{^t_t_{\rm, hydro}}, \tensor{T}{^t_i_{\rm, hydro}}\}$ where $\tensor{T}{^t_\mu_{\rm , hydro}}=\left(\rho+u+p\right)u^{t}u_{\mu}+p\tensor{\delta}{^t_\mu}$. Here, the coarse-grained conservative hydrodynamic variables are calculated by
\begin{equation}
    U_\mathrm{hydro,coarse}=\frac{\sum U_\mathrm{hdyro,fine}\Delta V_\mathrm{fine}}{\Delta V_\mathrm{coarse}},
\end{equation}
where $U_\mathrm{hydro, fine}$ is the variables in level $i-1$ being restricted and $\Delta V$ is the volume of the cells on the fine and coarse mesh. We always have $\Delta V_\mathrm{coarse}=8\Delta V_\mathrm{fine}$ for the Cartesian mesh we use. In this way, we conserve the hydrodynamic part of the conserved variables.

For the magnetic part, simply holding the magnetic field fixed in the mask region and enforcing the divergence-free constraint may cause a strong shear of the magnetic field lines around the boundary in the presence of rotation and inhomogeneity. Therefore, we still evolve the magnetic field $\boldsymbol{B}$ according to the induction equation using constrained transport to enforce the divergence-free constraint $\div\boldsymbol{B}=0$ as in typical MHD codes. Moreover, to preserve the Poynting flux, we add an extra source term from smaller scales in the mask region, i.e,
\begin{equation}
    \partial_t(\sqrt{-g} \boldsymbol{B}) + \nabla\times[\sqrt{-g} (\boldsymbol{\mathcal{E}} + \delta\boldsymbol{\mathcal{E}})] = 0.
    \label{eq:induction_delta_emf}
\end{equation}
Here, $\boldsymbol{\mathcal{E}}=\boldsymbol{B}\times\boldsymbol{u}_0$ is the usual electromotive force (EMF) except that the velocity $\boldsymbol{u}_0$ in the mask region is fixed. The extra term $\delta\boldsymbol{\mathcal{E}}$ is an additional constant EMF defined along the cell edges, the same way as the edge-centered EMF in \athenak{} and \athenapp~\citep{Stone2020ApJS..249....4S}. Instead of any ad hoc models, it is calculated along each direction by
\begin{equation}
    \delta\mathcal{E}=\hat{\mathcal{E}}_0-\mathcal{E}_0,
\end{equation}
where
\begin{equation}
    \hat{\mathcal{E}}_0=\frac{\sum \mathcal{E}_\mathrm{fine}\Delta L_\mathrm{fine}}{\Delta L_\mathrm{coarse}},
\end{equation}
is the average of the edge-centered EMF from the previous finer level right before the mesh derefinement and $\mathcal{E}_0$ is the EMF computed right after the mesh derefinement. Here, $\Delta L$ is the length of the cell edges and we always have $\Delta L_\mathrm{coarse}=2\Delta L_\mathrm{fine}$ in the Cartesian grid we use. The extra source term is thus the difference between the coarse-grained EMF from the finer variables and the EMF from the coarse-grained variables. In addition, we limit the source term such that $|\delta\mathcal{E}|\leq C\max(\mathcal{E}_0)$ in the mask region to keep the stability of the simulation. Here, the prefactor of upper limit $C$ is typically set to be $C=1$. The source term vanishes if the fluid is uniform. We find that inclusion of $\delta \boldsymbol{\mathcal{E}}$ is important for preserving the outgoing Poynting flux in the case of rotating black holes.

The conserved variables inside the mask region are thus computed using the fixed primitive variables $W_0$ and the evolved magnetic field $\boldsymbol{B}$ correspondingly instead of evolving according to the GRMHD equations. This means that the code is no longer conservative to machine precision in the mask region. But this is expected since the code effectively ``skips'' evolution of the fluid within $r_{\mathrm{z}}$. We have numerically verified that the mass change due to this algorithm is approximately equal to that in an analogous normal GRMHD run. The relative difference is $\lesssim 1\%$ (see Appendix~\ref{app:cons} for details).

Correspondingly, when we zoom in from level $i+1$ to level $i$ with mesh refinement, we perform prolongation for the evolved coarse variables outside the old mask region $(r > r_{\mathrm{z},i+1})$. But for variables within the old mask region $(r < r_{\mathrm{z},i+1})$, since the hydrodynamic variables are not evolved and remain unchanged, we set the primitive variables to be the previously stored finer primitive variables on level $i$ as the best guess instead of prolongation. This will inevitably bring the two states from different physical times into contact at the old mask region boundary ($r_{\mathrm{z},i+1}$), which will cause discontinuities if the system is inhomogeneous. But using anything else will invariably lose crucial information about the small scales and thus introduce extra errors larger than the current implementation. We still perform prolongation for the magnetic field within the mask region, since it has evolved. Then, for variables within the new mask region $(r < r_{\mathrm{z},i})$, we still keep the primitive variables fixed during this level and apply the source term $\delta\boldsymbol{\mathcal{E}}$ previously stored on level $i$. Though not presented in this work, if we start the simulation using the ``V-cycle,'' we will fill the mask region with predefined initial conditions (e.g., a ``vacuum sink,'' a uniform medium, or a Bondi solution) in the first half cycle.

Finally, the runtime at each level $i$ is typically set by
\begin{equation}
    t_{\mathrm{run},i}=f t_{\mathrm{z},i},
\end{equation}
where $t_{\mathrm{z},i}$ is the characteristic timescale at $r_{\mathrm{z},i}$ which is defined more precisely below for each problem we study, depending on the characteristic speed at $r_{\mathrm{z},i}$. The parameter $f$ is a factor of $\sim O(10)$ to $\sim O(100)$ such that the runtime is sufficient for communication of information between scales. We have verified that the results do not change significantly as long as $f$ is within this range. We typically skip the initial few levels $i=1$ and $i=2$ to avoid strong GR effects around the boundary of the mask region, which otherwise would make the mass and energy flux from the horizon scale that is preserved in the mask region less accurate. At level $i=1,\,2$, we are still partially resolving the horizon with $r_\mathrm{g}/\Delta x = 8,\,4$, respectively. Then the solution is a combination of the poorly resolved horizon rotation around the ergosphere and the extra source term, which may easily be biased. Therefore, we prefer that the horizon is either completely well resolved or not resolved at all. The downside of this approach is that we have to ensure that the runtime at $i=0$ is long enough that the solution converges within $8\,r_\mathrm{g}$ or further. We shall list the values of runtime parameters for each run below.

As an illustration of the cyclic zoom method in an actual simulation, \figu\ref{fig:method} (bottom) shows an example of $\Lambda$-cycle for accretion of a Fishbone-Moncfrief \citep[FM;][]{Fishbone1976ApJ...207..962F} torus in magnetically arrested state onto a black hole with spin $a=0.9375$ when the system is quasi-steady. The mask region can keep the small-scale structure and provide correct feedback from the central region. We will present a detailed analysis for the torus accretion in \sect\ref{subsec:valid_torus}.

We note that the goal of the cyclic zoom approach is to obtain statistically equivalent outcomes when turbulent situations are considered, with qualitatively similar flow features. Identical outcomes are of course not obtained and are usually also not needed. In fact, even the conventional fluid simulations involving turbulence can show different outcomes after a long duration, due to floating-point roundoff errors, if we rerun it.

The speedup of the method is shown by the computation time plot in \figu\ref{fig:method} (top right) measured for the case of magnetized Bondi accretion onto a black hole (see \sects\ref{subsec:valid_bondi} and \ref{subsec:app_bondi} for details of the simulation setup). Typical GRMHD simulations follow a scaling of $\propto r^{3/2}$, which quickly becomes prohibitive for growing characteristic radius (Bondi radius for Bondi accretion). Note that this is difficult to accelerate by increasing computational resources (number of CPUs/GPUs), but limited by the time-step constraints. Instead, cyclic zoom runs are relatively insensitive to the characteristic radius. Therefore, this method can accelerate the GRMHD simulations by a factor of $\sim 10^5$ or more when the characteristic radius is large ($\sim 10^6\,r_\mathrm{g}$). We will discuss the details of speedup in~\sect\ref{subsec:speedup}.

In summary, the cyclic zoom framework effectively uses the mask region as a ``subgrid'' model. However, instead of ad hoc analytic or semi-analytic assumptions of the subgrid model, we use the information directly from a full GRMHD simulation to provide the ``inner boundary.'' The cyclic zoom method can be thought of as an analog of AMR in time space, i.e., ``adaptive time refinement'' (not to be confused with adaptive time stepping, which updates different cells with different frequencies according to the Courant constraint). It can also be viewed as sampling the small-scale physics in the spacetime.

\subsection{Diagnostics}
Following previous GRMHD simulations, we describe the diagnostics used in this work. The relevant diagnostics include mass flux, magnetic field flux, energy flux, and angular momentum flux:
\begin{align}
    \dot{M}&\equiv-\int_S \rho u^r\sqrt{-g}\,d \Omega,\\
    \Phi_\mathrm{BH}&\equiv \sqrt{\pi} \int_S |B^r|\sqrt{-g}\,d\Omega,\\
    \dot{E}&\equiv \int_S \tensor{T}{^r_t}\sqrt{-g}\,d\Omega,\\
    \dot{L}&\equiv \int_S \tensor{T}{^r_\phi}\sqrt{-g}\,d\Omega,
\end{align}
where $S$ is the area, $\tensor{T}{^r_t}=(\rho+u+p+b^2)u^{r}u_{t}-\tensor{b}{^r}\tensor{b}{_t}$, and $\tensor{T}{^r_\phi}=(\rho+u+p+b^2)u^{r}u_{\phi}-\tensor{b}{^r}\tensor{b}{_\phi}$.
Sometimes it is useful to separate the mass flux into inflow and outflow:
\begin{equation}
    \dot{M}=\underbrace{-\int_{u^r<0}\rho u^r\sqrt{-g}\,d \Omega}_{\dot{M}_\mathrm{in}}-\underbrace{\int_{u^r>0}\rho u^r\sqrt{-g}\,d \Omega}_{\dot{M}_\mathrm{out}}.
\end{equation}
The dimensionless magnetic flux parameter is defined by
\begin{equation}
    \phi_\mathrm{BH}\equiv\frac{\Phi_\mathrm{BH}}{\sqrt{\dot{M}}}.
\end{equation}
The energy flux $\dot{E}$ includes the flux of rest-mass energy, so the ``feedback'' energy flux is $\dot{M}-\dot{E}$. The efficiency of feedback can thus be defined by
\begin{equation}
    \eta\equiv \frac{\dot{M}-\dot{E}}{\dot{M}}.
\end{equation}
The efficiency is positive (negative) when energy is transported outward (inward). We can further separate the feedback power into the hydrodynamic part 
\begin{equation}
    \dot{E}_\mathrm{hydro}=-\int_S \left[(\rho+u+p)u^{r}u_{t}+\rho u^r\right]\sqrt{-g}\,d\Omega,
\end{equation}
and the electromagnetic (EM) part
\begin{equation}
    \dot{E}_\mathrm{EM}=-\int_S \left(b^2u^{r}u_{t}-\tensor{b}{^r}\tensor{b}{_t}\right)\sqrt{-g}\,d\Omega,
\end{equation}
Finally, the specific angular momentum flux is defined by
$l\equiv {\dot{L}}/ {\dot{M}}$.

\setlength{\tabcolsep}{4pt}
\begin{deluxetable*}{l|c|cccccccccc}
    \tablenum{1}
    \tablecaption{List of standard, validation, and application runs and their main parameters in this work. \label{tab:runs}}
    \tablewidth{0pt}
    \tablehead{
    \colhead{Run} &\colhead{Model} & \colhead{Spin $a$} & \colhead{$r_\mathrm{B}$} & \colhead{$\beta$} & \colhead{$r_\mathrm{in}$} & \colhead{$r_\mathrm{out}$} & \colhead{$r_\mathrm{peak}$} & \colhead{$\gamma_\mathrm{ad}$} & Half Box & AMR &  \colhead{Runtime $t_\mathrm{run}$} 
    \\[-0.2cm] \colhead{Type} & \colhead{Type} & \colhead{} & \colhead{$(r_\mathrm{g})$} & \colhead{} & \colhead{$(r_\mathrm{g})$} & \colhead{$(r_\mathrm{g})$} & \colhead{$(r_\mathrm{g})$} & & Size $(r_\mathrm{g})$ & Level & \colhead{$(t_\mathrm{g})$}
    }
    \startdata
    \multirow{13}{*}{
    \begin{tabular}{c}
    Standard / \\
    Validation \\
    (\S\ref{sec:validation})
    \end{tabular}
    } & \begin{tabular}{c}
    Hydro \\
    Bondi \\
    (\S\ref{subsec:valid_hyd_bondi})
    \end{tabular} & $0$ & $\approx800$ & & & & & $5/3$ & $2^{17}\approx10^5$ & $15$ & $10^5$ ($\approx 6 t_\mathrm{B}$)
    \\ \cline{2-12} & \begin{tabular}{c}
    Monopole \\
    (\S\ref{subsec:valid_monopole})
    \end{tabular} & $0$ & & & & & & $4/3$ & $2^{14}\approx10^4$ & $12$ & $4\times10^3$
    \\ \cline{2-12} & \multirow{4}{*}{
    \begin{tabular}{l}
    MHD \\
    Bondi \\
    (\S\ref{subsec:valid_bondi})
    \end{tabular}
    } & $0$ & $500$ & $1$ & & & & $5/3$ & $2^{17}\approx10^5$ & $15$ & $2\times10^5$ ($\approx 25 t_\mathrm{B}$)
    \\ &  & $0.9375$ & $500$ & $1$ & & & & $5/3$ & $2^{17}\approx10^5$ & $15$ & $2\times10^5$ ($\approx 25 t_\mathrm{B}$)
    \\ &  & $0$ & $500$ & $10^3$ & & & & $5/3$ & $2^{17}\approx10^5$ & $15$ & $2\times10^5$ ($\approx 25 t_\mathrm{B}$)
    \\ &  & $0.9375$ & $500$ & $10^3$ & & & & $5/3$ & $2^{17}\approx10^5$ & $15$ & $2\times10^5$ ($\approx 25 t_\mathrm{B}$)
    \\ \cline{2-12} & \multirow{5}{*}{
    \begin{tabular}{l}
    Torus \\
    (\S\ref{subsec:valid_torus})
    \end{tabular}
    } & $-0.9375$ &  & & $20$ & $10^4$ & $43.08$ & $13/9$ & $2^{17}\approx10^5$ & $15$ & $10^5$
    \\ &  & $-0.5$ &  & & $20$ & $10^4$ & $42.75$ & $13/9$ & $2^{17}\approx10^5$ & $15$ & $10^5$
    \\ &  & $0$ &  & & $20$ & $10^4$ & $42.42$ & $13/9$ & $2^{17}\approx10^5$ & $15$ & $10^5$
    \\ &  & $0.5$ &  & & $20$ & $10^4$ & $42.14$ & $13/9$ & $2^{17}\approx10^5$ & $15$ & $10^5$
    \\ &  & $0.9375$ &  & & $20$ & $10^4$ & $41.94$ & $13/9$ & $2^{17}\approx10^5$ & $15$ & $10^5$
    \\ \cline{2-12} & \multirow{2}{*}{
    \begin{tabular}{l}
    Torus \\
    (\S\ref{subsec:valid_secular})
    \end{tabular}
    } & $0$ &  & & $20$ & $100$ & $34.49$ & $13/9$ & $2^{17}\approx10^5$ & $15$ & $10^6$
    \\ &  & $0.9375$ &  & & $20$ & $100$ & $34.23$ & $13/9$ & $2^{17}\approx10^5$ & $15$ & $2\times10^5$
    \\\hline
     \multirow{6}{*}{
     \begin{tabular}{c}
     \\ \\
     Application \\
     (\S\ref{sec:results})
     \end{tabular}
     }  & \multirow{4}{*}{
    \begin{tabular}{l}
    MHD \\
    Bondi \\
    (\S\ref{subsec:app_bondi})
    \end{tabular}
    } & $0$ & $2\times10^4$ & $1$ & & & & $5/3$ & $2^{20}\approx10^6$ & $18$ & $3\times10^7$ ($\approx 15 t_\mathrm{B}$)
    \\ &  & $0.9375$ & $2\times10^4$ & $1$ & & & & $5/3$ & $2^{20}\approx10^6$ & $18$ & $3\times10^7$ ($\approx 15 t_\mathrm{B}$)
    \\ &  & $0$ & $4\times10^5$ & $1$ & & & & $5/3$ & $2^{24}\approx10^7$ & $22$ & $10^9$ ($\approx 5 t_\mathrm{B}$)
    \\ &  & $0.9375$ & $4\times10^5$ & $1$ & & & & $5/3$ & $2^{24}\approx10^7$ & $22$ & $10^9$ ($\approx 5 t_\mathrm{B}$)
    \\ \cline{2-12} & \begin{tabular}{c}
    Torus \\
    (\S\ref{subsec:app_torus})
    \end{tabular} & $0.9375$ &  & & $10^3$ & $10^4$ & $1820$ & $13/9$ & $2^{24}\approx10^7$ & $22$ & $10^7$ 
    \\ \cline{2-12} & \begin{tabular}{c}
    Torus \\
    (\S\ref{subsec:app_secular})
    \end{tabular} & $0$ &  & & $20$ & $100$ & $34.49$ & $13/9$ & $2^{17}\approx10^5$ & $15$ & $10^7$ 
    \enddata
    \tablecomments{The validation and the standard runs are performed with the same spatial resolution over the same time duration. All runs are performed with a finest resolution of $\Delta x_\mathrm{min} = r_\mathrm{g}/16$ and a relative resolution on a grid of $128^3$ cells per refinement level, except for the monopole test, which only covers the domain of $z>0$ using $128\times128\times64$ cells per level. The number of AMR levels is adjusted according to the box size.}
\end{deluxetable*}

\begin{figure*}[ht!]
\includegraphics[width=\linewidth]{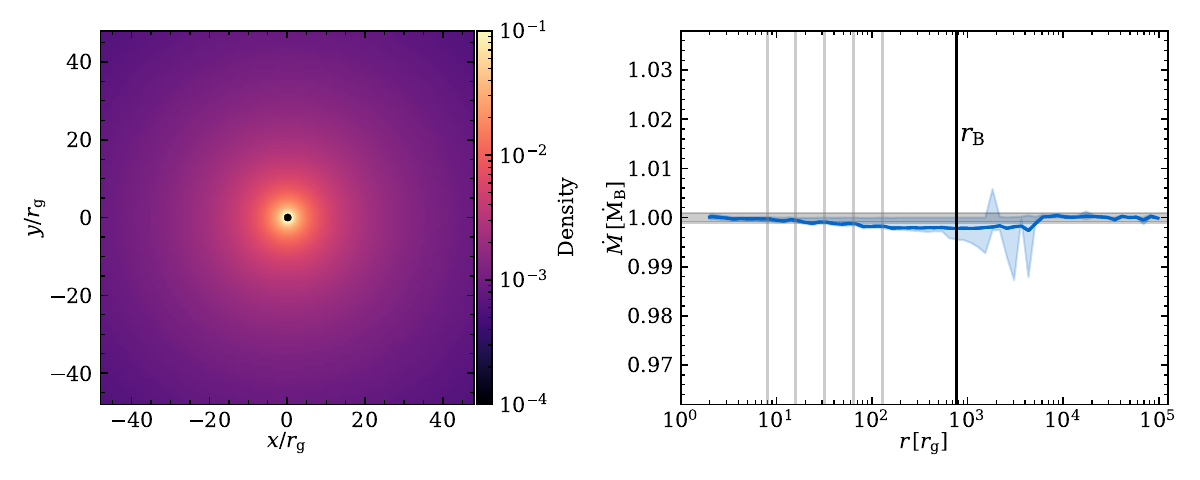}
\caption{Left: Logarithmic mid-plane density $\rho$ ($z=0$ slice) from a spherically symmetric Bondi accretion with a Bondi radius of $\sim800\,r_\mathrm{g}$ onto a Schwarzschild black hole at the end of the simulation ($t\approx10^5\,t_\mathrm{g}\approx6\,t_\mathrm{B}$) using the cyclic zoom method. Right: radial profile of time-averaged accretion rate with shaded range showing the 10\%-90\% time inclusion interval. The vertical black line marks the Bondi radius $r_\mathrm{B}$ and the vertical light gray lines mark the boundary of the mask region for all levels. The gray shaded region marks $\pm0.1\%$ difference from the analytic Bondi accretion rate. The cyclic zoom method maintains the analytic solution within $\sim 0.1\%$ over $\gtrsim 6$ Bondi time.
\label{fig:hyd_bondi}}
\end{figure*}

\begin{figure*}[ht!]
\includegraphics[width=\linewidth]{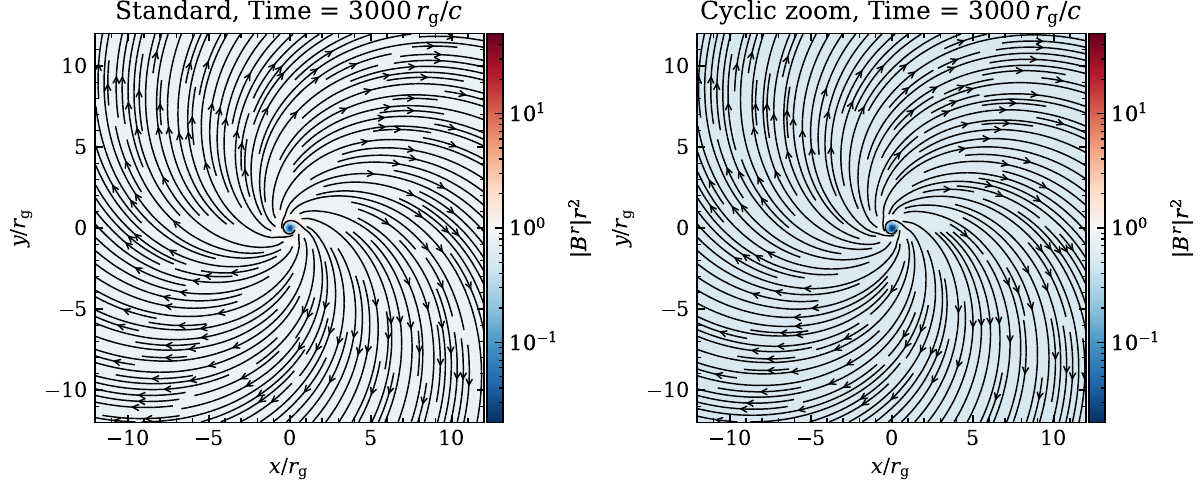}
\includegraphics[width=0.49\linewidth]{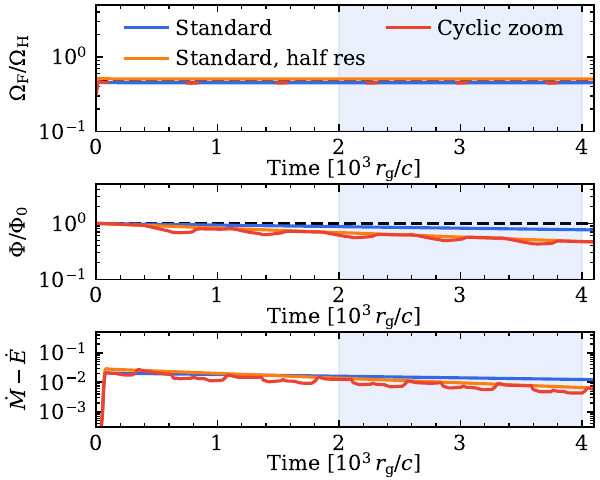}
\includegraphics[width=0.49\linewidth]{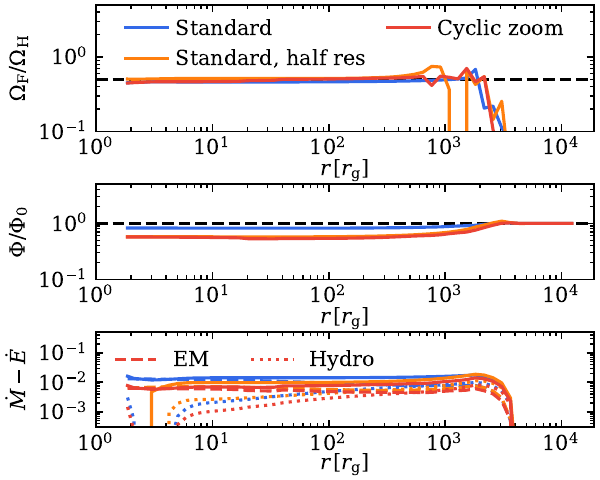}
\caption{Top: The $z=0$ slice of magnetic field flux $|B^r|r^2$ with the magnetic field lines from a monopole around a Kerr black hole with $a=0.5$ in a standard run (left) and a cyclic zoom run (right). The rotational symmetry is preserved even after $3000\,t_\mathrm{g}$ in the cyclic zoom run. Bottom left: history of field rotation rate measured at the horizon, the magnetic flux at $r=48\,r_\mathrm{g}$ and feedback power at $r=48\,r_\mathrm{g}$ for the standard run, the cyclic zoom run, and a standard run with half of the resolution everywhere. We average values over the second half of the simulations (blue background) for time-averaged properties. The field rotation rate is recovered accurately. The magnetic flux and feedback power are roughly constant with slow decay, similar to those in the standard run with half resolution. Bottom right: radial profile of time-averaged field rotation rate, magnetic flux, and feedback power. The cyclic zoom method can reproduce the field rotation rate and the energy flux.
\label{fig:mono_valid}}
\end{figure*}

\begin{figure*}[ht!]
\includegraphics[width=\linewidth]{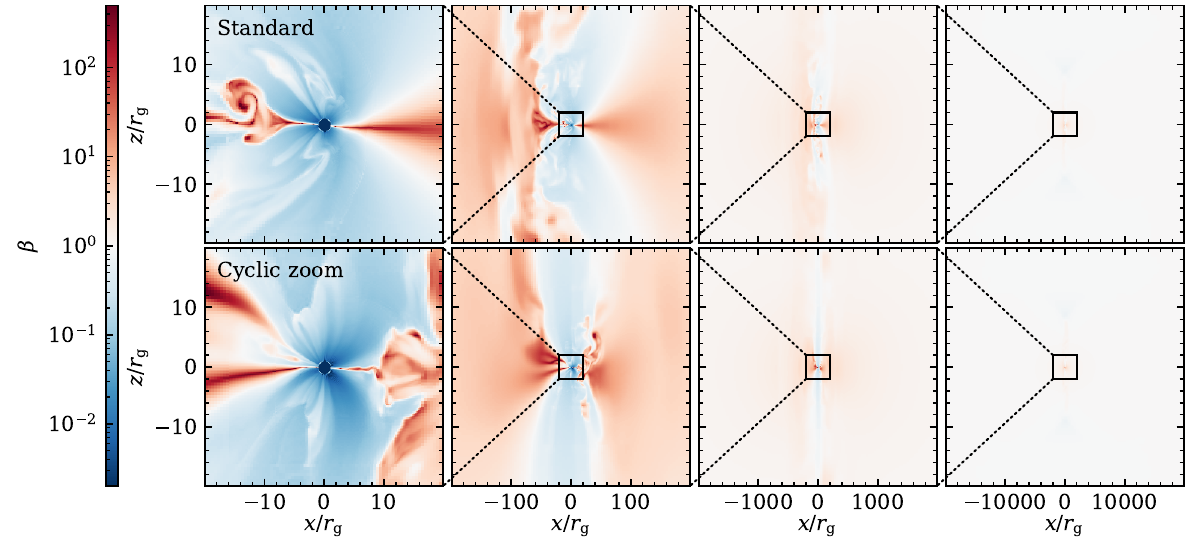}
\includegraphics[width=\linewidth]{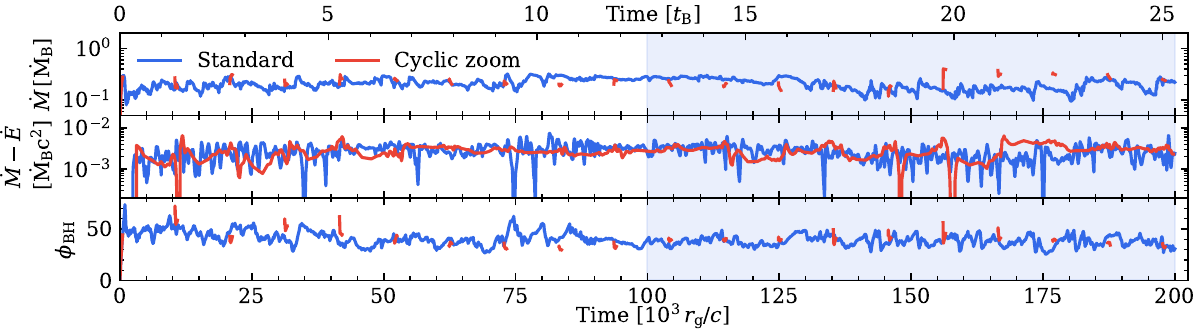}
\includegraphics[width=\linewidth]{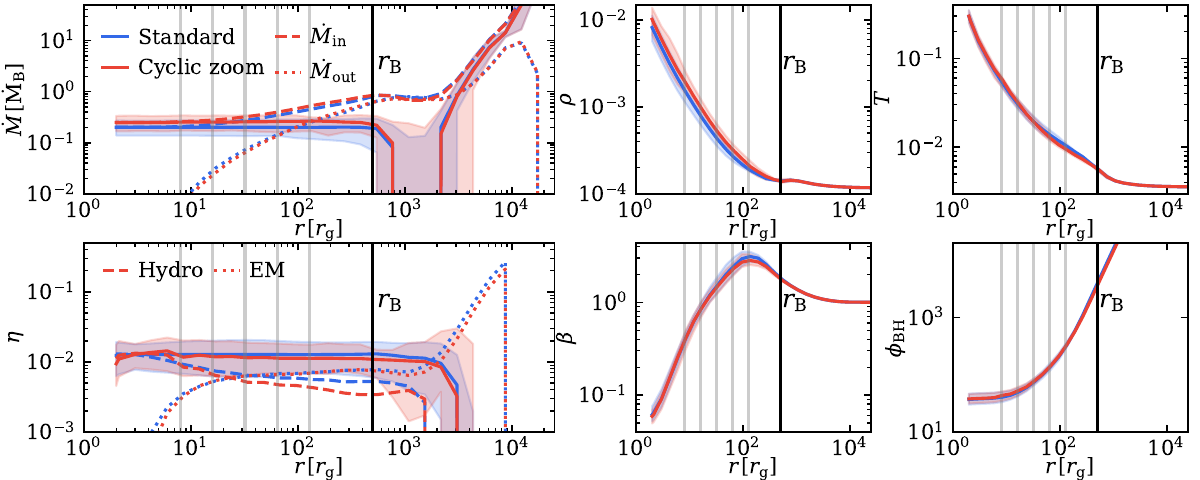}
\caption{Comparison of the cyclic zoom run with the standard run of Bondi accretion of magnetized plasma with Bondi radius $r_\mathrm{B}=500\,r_\mathrm{g}$ and initial $\beta=1$ onto a black hole with spin $a=0$. Top: the $y=0$ slice of plasma-$\beta$ on different scales in the standard run (upper) and the cyclic zoom run (lower) at nearly the end of the simulation. Middle: smoothed history of accretion rate measured at $r=3\, r_\mathrm{g}$ normalized by the Bondi accretion rate, energy feedback rate measured at $r=200\, r_\mathrm{g}$, and the magnetic flux parameter measured at $r=3\, r_\mathrm{g}$. We average values over the second half of the simulations (blue background) for time-averaged properties. Bottom: time-averaged radial profiles of accretion rate with inflow and outflow, density, temperature, feedback efficiency including hydrodynamic part and EM part, plasma-$\beta$, and magnetic flux parameter. The vertical black line marks the Bondi radius $r_\mathrm{B}$ and the vertical light gray lines mark the boundary of the mask region for all levels. The shaded ranges show the 10\%-90\% time inclusion interval. The cyclic zoom method captures the evolution, averaged properties, and variability of the accretion flow.
\label{fig:bondi_valid}}
\end{figure*}

\begin{figure*}[ht!]
\includegraphics[width=\linewidth]{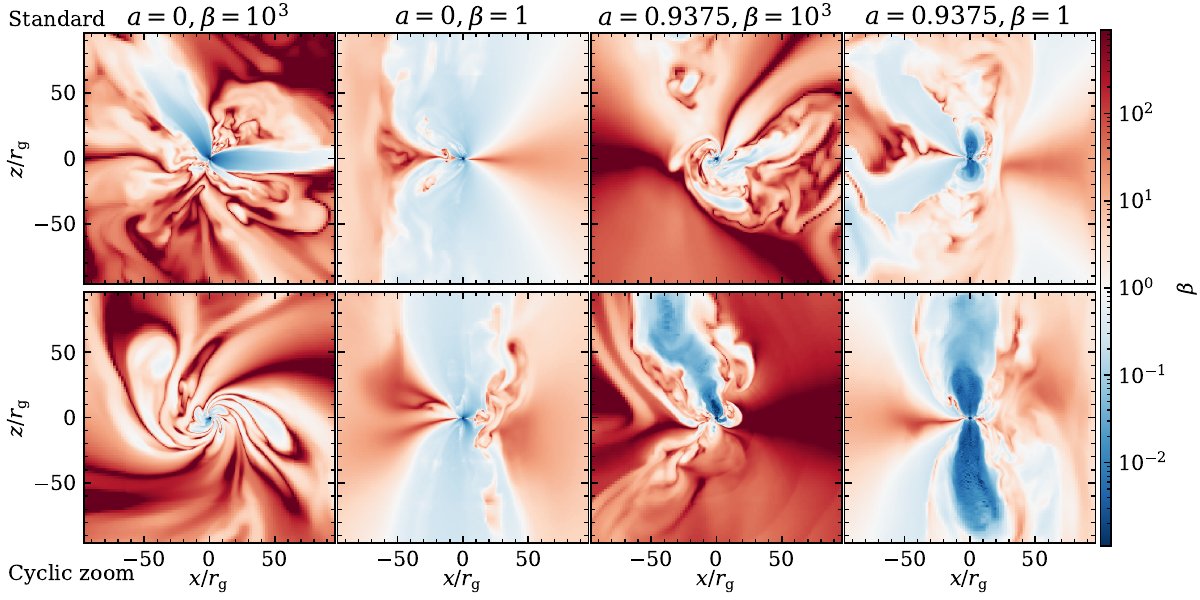}
\includegraphics[width=\linewidth]{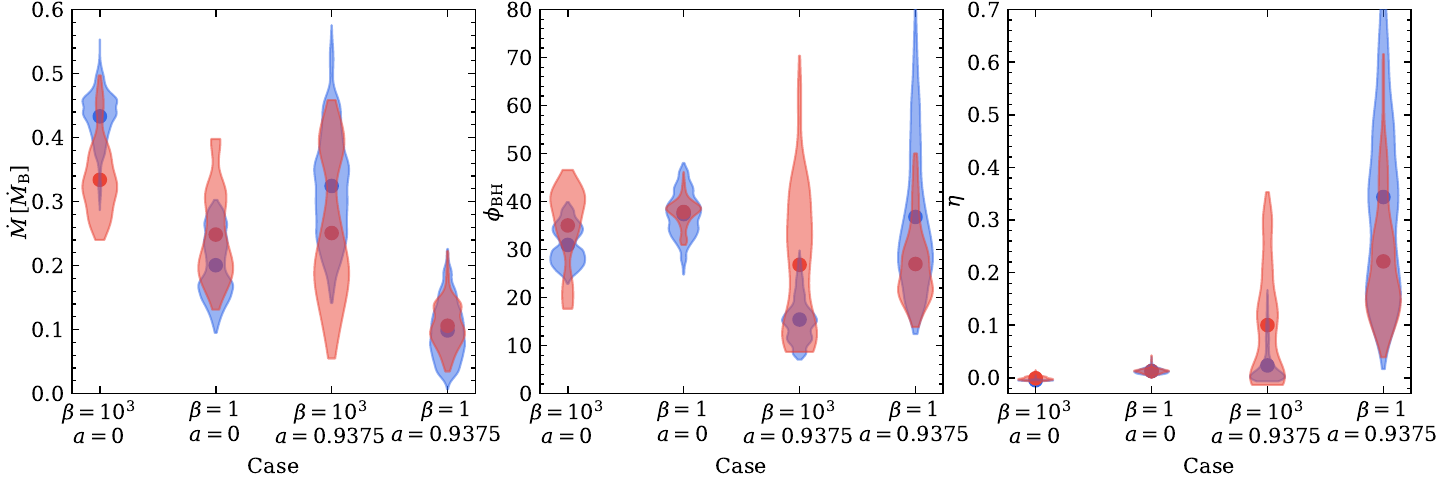}
\caption{Top: The $y=0$ slice of plasma-$\beta$ in the standard runs (top) and the cyclic zoom runs (bottom) in different cases at nearly the end of the simulations. The cyclic zoom runs show behaviors similar to those of the standard runs. Bottom: distribution of accretion rate (left), magnetic flux parameter (middle), and feedback efficiency (right) of the cyclic zoom method (red) with the standard (blue) runs for Bondi accretion with spin $a=0,0.9375$ and plasma-$\beta=1000,1$. Compared with the standard runs, the cyclic zoom runs produce similar mean accretion rate, magnetic flux, and feedback efficiency with similar dispersion.
\label{fig:bondi_valids}}
\end{figure*}

\begin{figure*}[ht!]
\includegraphics[width=\linewidth]{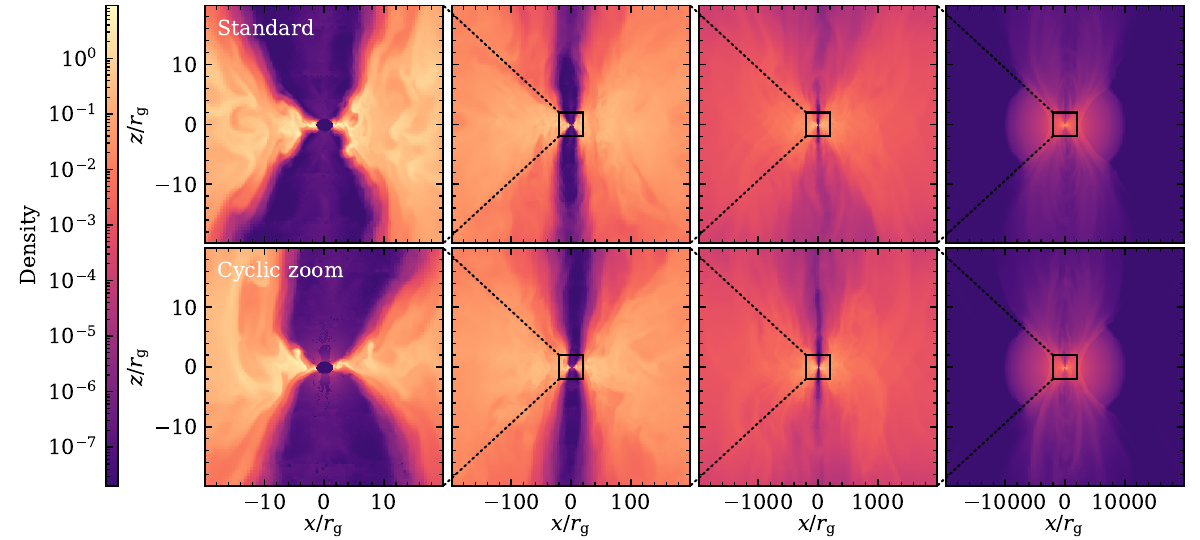}
\includegraphics[width=\linewidth]{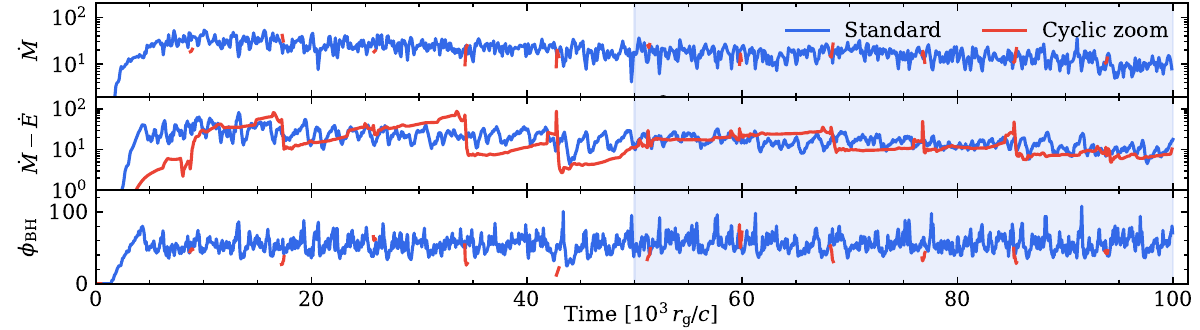}
\includegraphics[width=\linewidth]{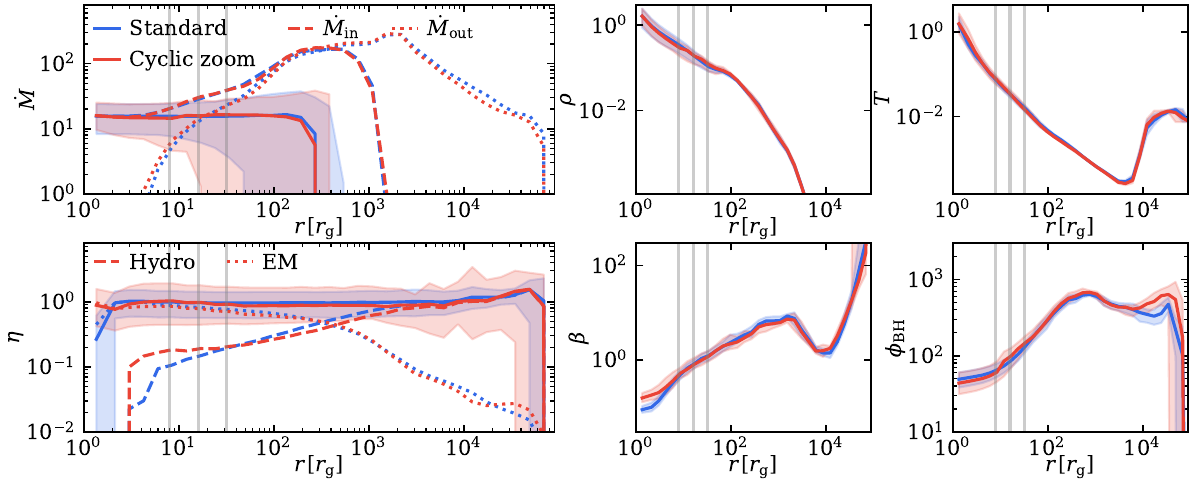}
\caption{Similar to \figu\ref{fig:bondi_valid}, but for accretion of an FM torus with initial $r_\mathrm{in}= 20\,r_\mathrm{g}$ and $r_\mathrm{out}= 10^4\,r_\mathrm{g}$ onto a spinning black with $a=0.9375$. The cyclic zoom method can reproduce the feedback efficiency of $\eta\sim100\%$ shown in the standard run even if the EM part dominates the energy flux.
\label{fig:torus_valid}}
\end{figure*}

\begin{figure*}[ht!]
\includegraphics[width=\linewidth]{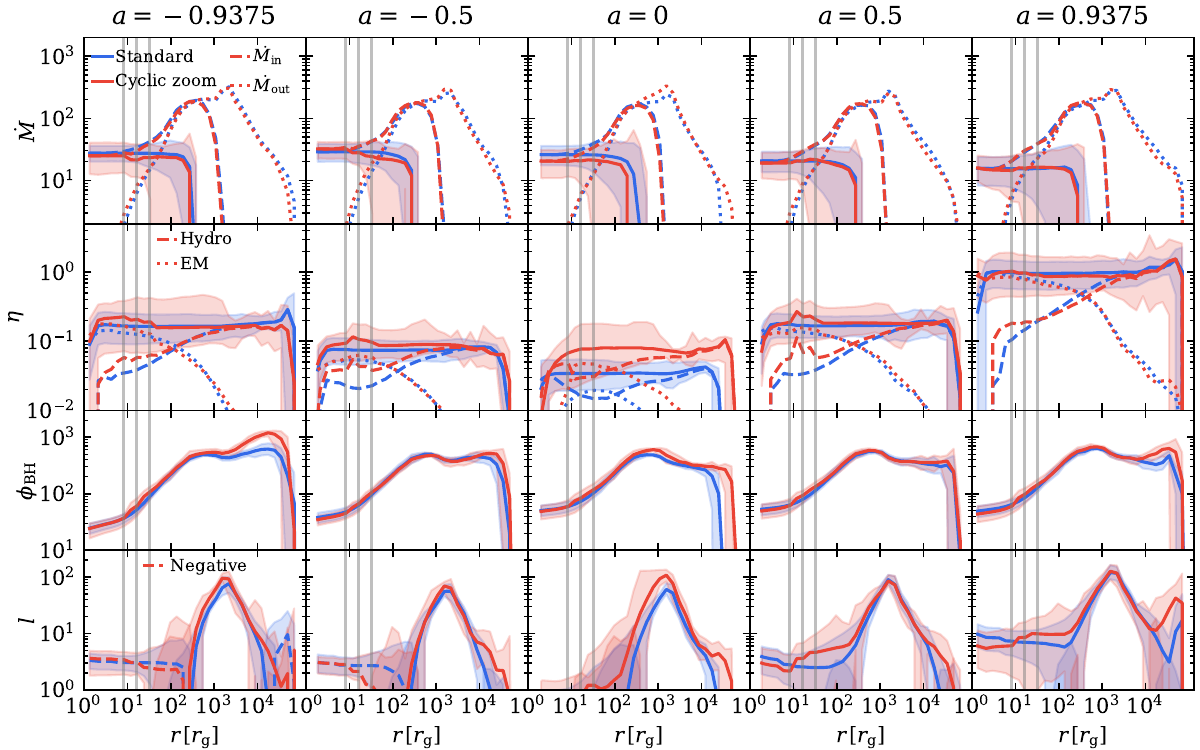}
\includegraphics[width=\linewidth]{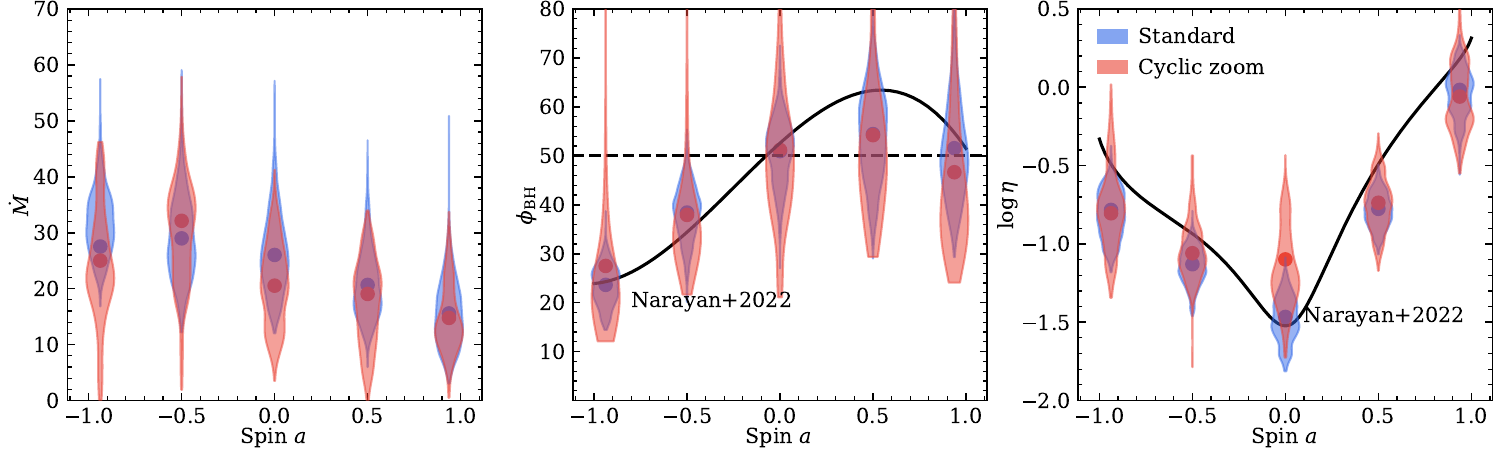}
\caption{
Comparison of the cyclic zoom method (red) with the standard (blue) runs for torus accretion with various spins. Top: time-averaged radial profiles of (from top to bottom) accretion rate, feedback efficiency, magnetic flux parameter, and specific angular momentum flux for different spins. The shaded region shows the $10\%-90\%$ time inclusion interval. The cyclic zoom runs show behaviors similar to those of the standard runs. Bottom left: distribution of accretion rate as a function of spin. Bottom center: distribution of magnetic flux parameter with the black solid line being fit to $\phi_\mathrm{BH}(a)$ from \citet{Narayan2022MNRAS.511.3795N}. Bottom right: distribution of feedback efficiency. The black line is the fit from \citet{Narayan2022MNRAS.511.3795N} plus a constant efficiency of $3\%$ as an estimate of the feedback from the accretion disk itself. The cyclic zoom method can reproduce the results for all spins.
\label{fig:torus_valids}}
\end{figure*}

\begin{figure}[t!]
\includegraphics[width=\linewidth]{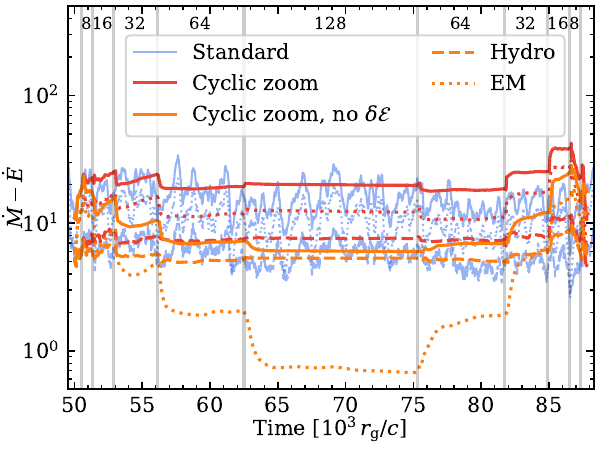}
\caption{Energy feedback rate measured at $r=135\,r_\mathrm{g}$ as a function of time including hydrodynamic and EM part for an FM torus around a black hole with $a=0.9375$. We show a standard run and two cyclic zoom runs with and without the source term $\delta \boldsymbol{\mathcal{E}}$ in \eq\ref{eq:induction_delta_emf}. The vertical lines mark the time for zooming out and zooming in with the corresponding $r_{\mathrm{z},i}$. The EM part for the cyclic zoom run without $\delta \boldsymbol{\mathcal{E}}$ decreases by $\sim30\%$ each time when zooming out. The source term is important in maintaining the outgoing Poynting flux.
\label{fig:torus_valid_emf}}
\end{figure}

\begin{figure*}[ht!]
\includegraphics[width=\linewidth]{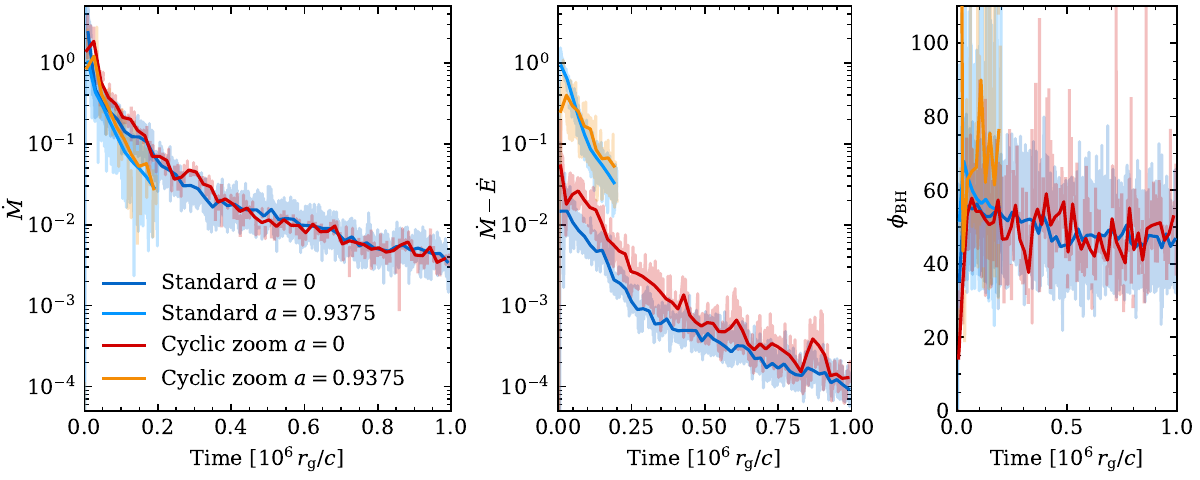}
\caption{Smoothed history of accretion rate at $r=3\, r_\mathrm{g}$, feedback energy flux at $r=200\, r_\mathrm{g}$, and magnetic flux parameter at $r=3\, r_\mathrm{g}$ for the cyclic zoom method and the standard runs of the long-term evolution of torus accretion with $a=0$ and $a=0.9375$. The shaded region shows the variability of the variables. The cyclic zoom method is capable of capturing the secular evolution of the viscous torus over a long time. \label{fig:secular_valid}}
\end{figure*}

\begin{figure*}[ht!]
\includegraphics[width=\linewidth]{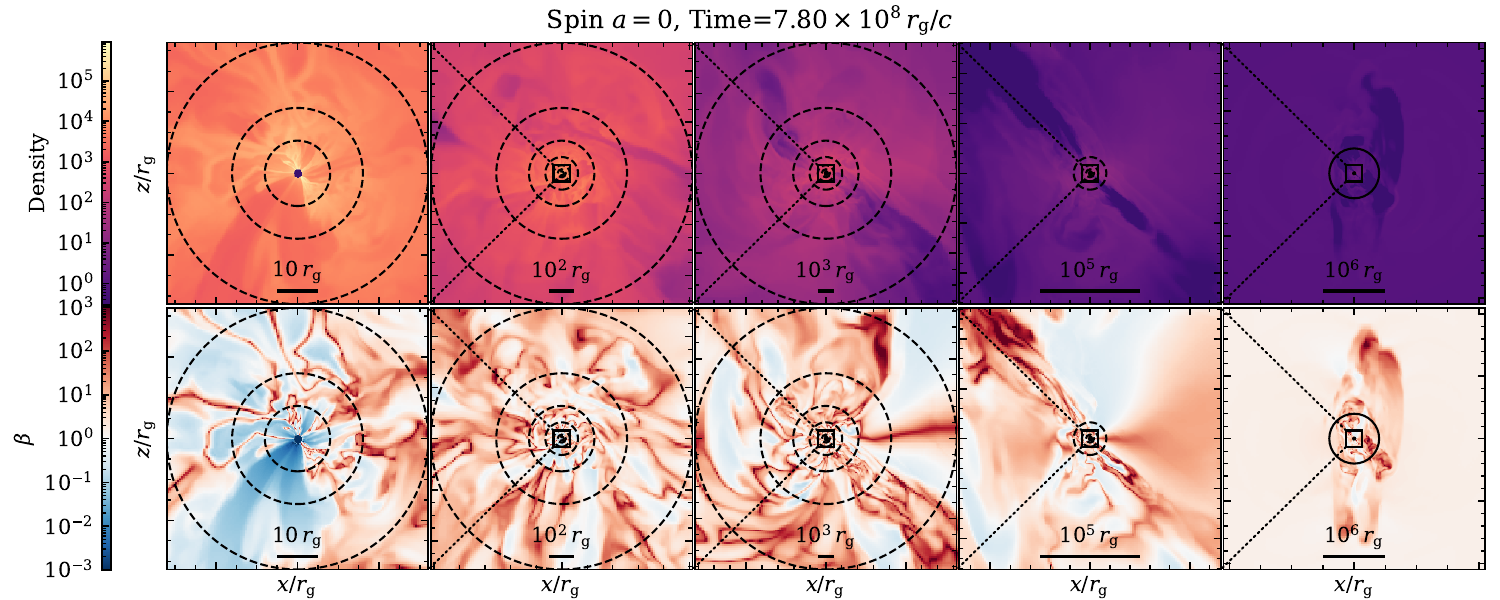}
\includegraphics[width=\linewidth]{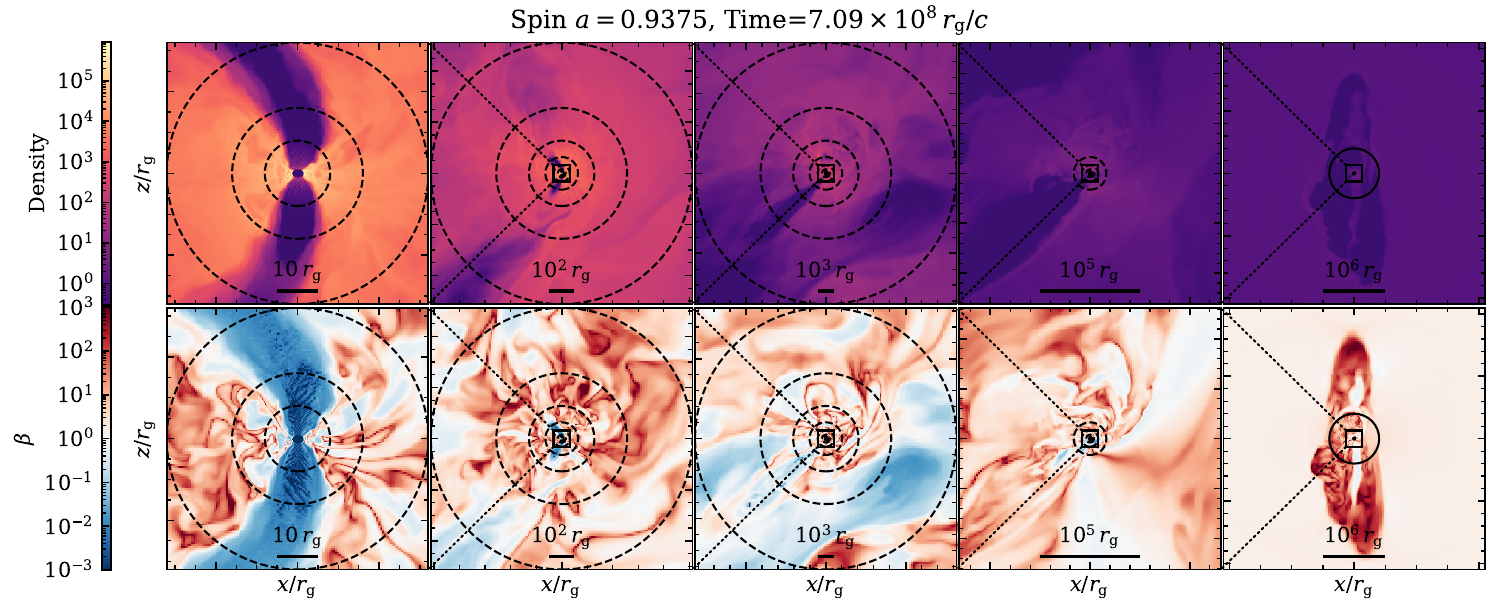}
\caption{Slice of $y=0$ plane in Bondi accretion with $r_\mathrm{B}=4\times10^5\,r_\mathrm{g}$ and spin $a=0$ (top) and $a=0.9375$ (bottom) using the cyclic zoom method. The dashed circles mark the boundary of each mask region. The solid circle in the rightmost panels marks the Bondi radius. The accretion flow is turbulent on all scales within the Bondi radius. The jet launched by the spinning black hole is typically choked well inside the Bondi radius.
\label{fig:bondi_slice}}
\end{figure*}

\begin{figure*}[ht!]
\includegraphics[width=\linewidth]{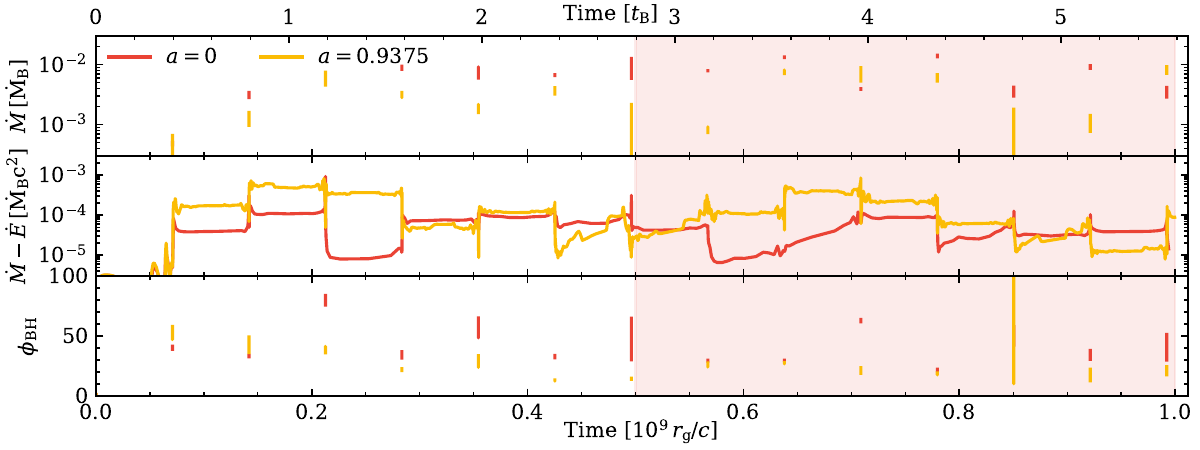}
\includegraphics[width=\linewidth]{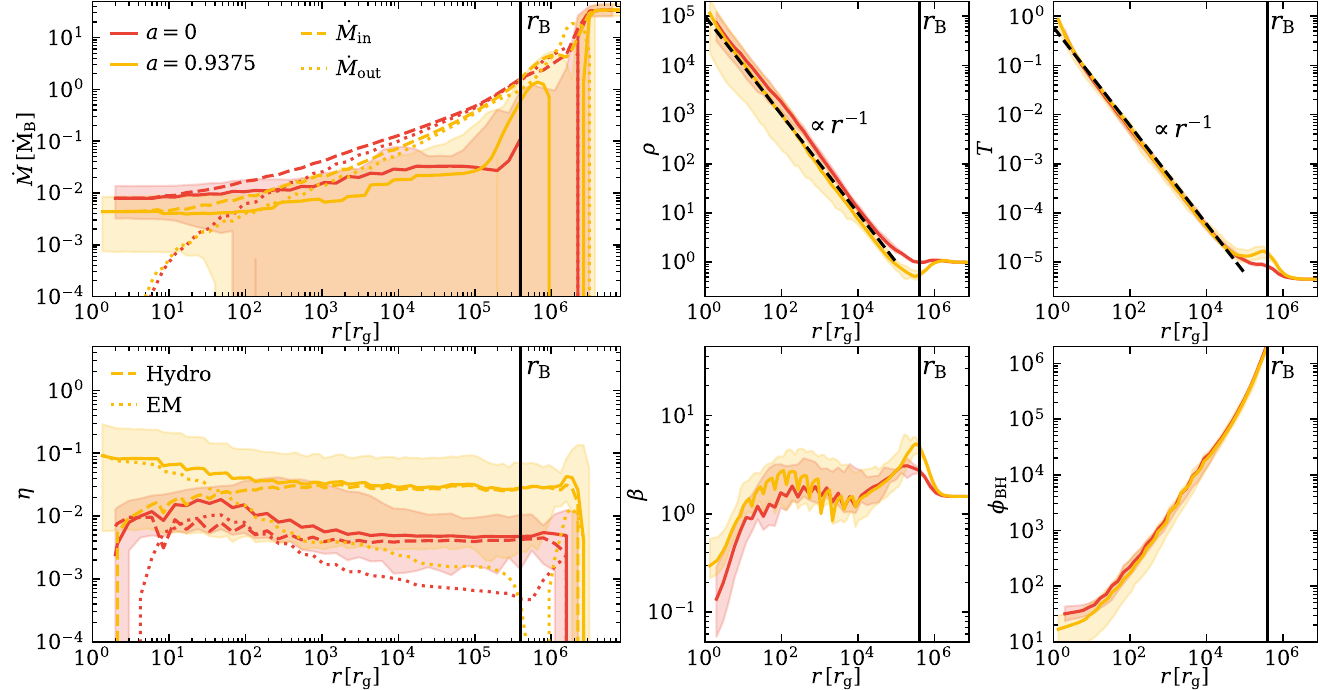}
\caption{Similar to \figu\ref{fig:bondi_valid} but for magnetized Bondi accretion with $r_\mathrm{B}=4\times10^5\,r_\mathrm{g}$ and $a=0$ (red) and $a=0.9375$ (yellow) using the cyclic zoom method. The history of feedback power is measured at $r\approx2\times10^4\,r_\mathrm{g}$. The time-averaged accretion rate is $1\%\dot{M}_\mathrm{B}$ and the feedback efficiency is $\sim 1\%$ for $a=0$ and $\sim 10\%$ for $a=0.9375$. The feedback efficiency has a wide distribution. The density scales with radius as $\rho\propto r^{-1}$.
\label{fig:bondi_application}}
\end{figure*}

\begin{figure}[ht!]
\includegraphics[width=\linewidth]{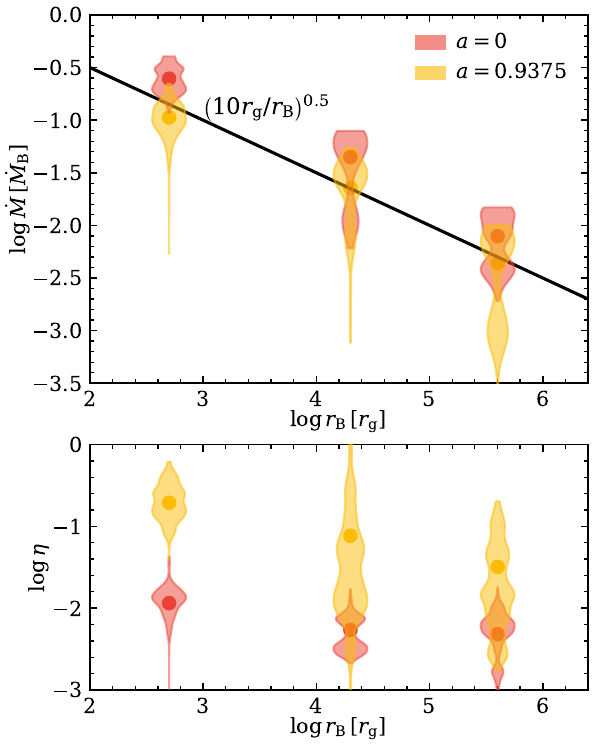}
\caption{Accretion rate measured at $r=3\, r_\mathrm{g}$ and feedback efficiency measured at $r\approx0.3\, r_\mathrm{B}$ as a function of Bondi radius for spin $a=0$ and $0.9375$. Accretion rate scales with Bondi radius as $\dot{M}_\mathrm{acc}/\dot{M}_\mathrm{B}\approx(10r_\mathrm{g}/r_\mathrm{B})^{1/2}$. The feedback efficiency is roughly $\sim 1\%$ for the Schwarzschild black hole and $\sim 10\%$ for the spinning black hole.
\label{fig:bondi_scaling}}
\end{figure}

\begin{figure*}[ht!]
\includegraphics[width=\linewidth]{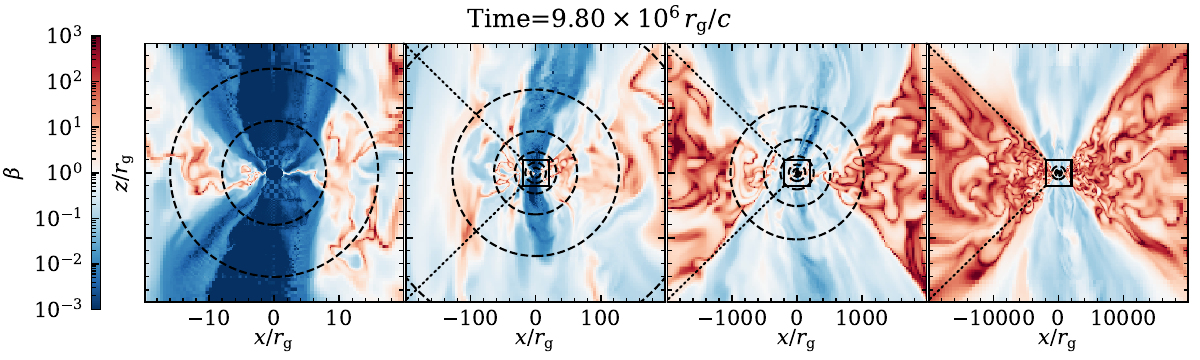}
\includegraphics[width=\linewidth]{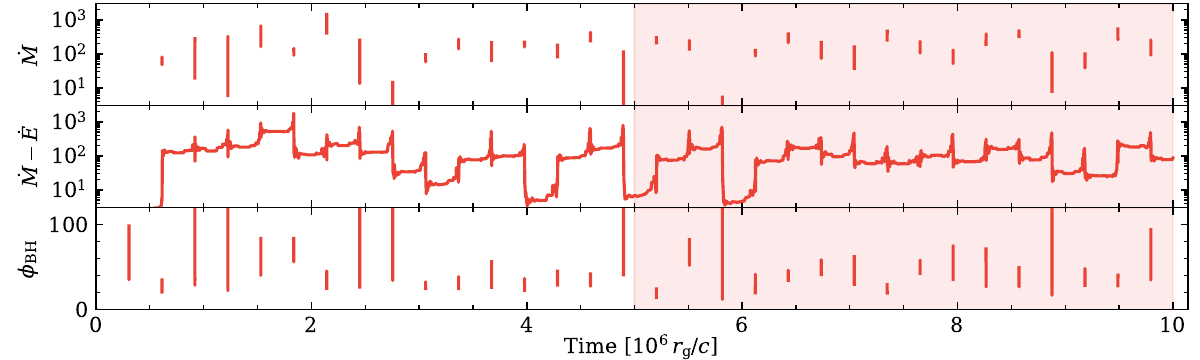}
\includegraphics[width=\linewidth]{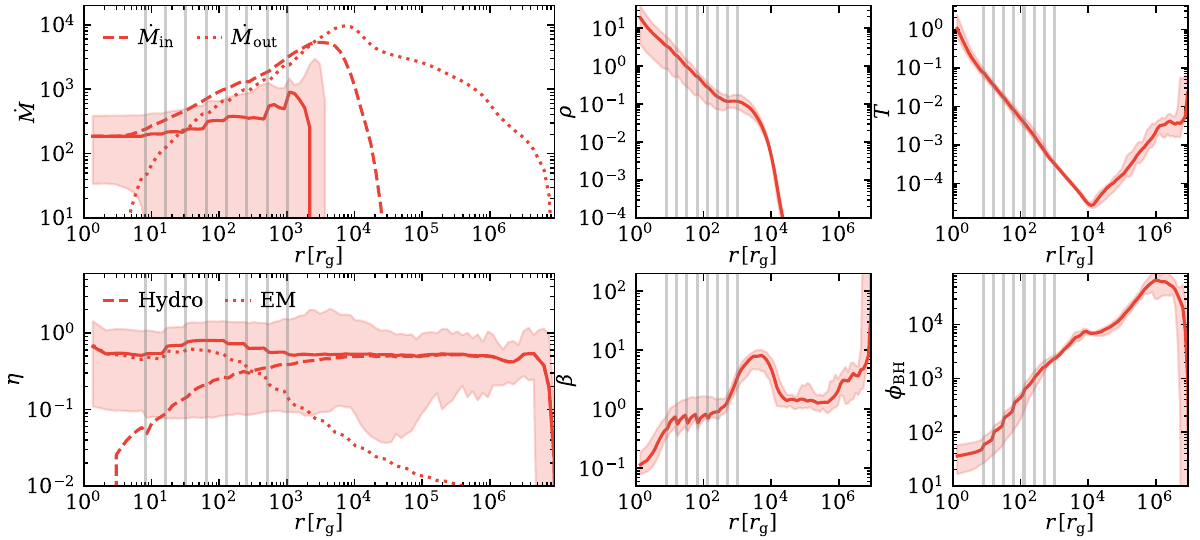}
\caption{Similar to \figu\ref{fig:bondi_valid} but for accretion of a large FM torus initially between $r_\mathrm{in}=10^3\,r_\mathrm{g}$ and $r_\mathrm{out}=10^4\,r_\mathrm{g}$ onto a spinning black hole with $a=0.9375$ using the cyclic zoom method. The dashed circles mark the boundary of each mask region. The history of feedback energy flux is measured at $r\approx10^3\,r_\mathrm{g}$. We reach a quasi-steady state within $\sim 2\times10^3\,r_\mathrm{g}$ and a constant feedback efficiency within $\sim 10^7\,r_\mathrm{g}$ with a mean value of $50\%$ fluctuation between $\sim 10\%-100\%$. 
\label{fig:torus_application}}
\end{figure*}

\begin{figure*}[ht!]
\includegraphics[width=\linewidth]{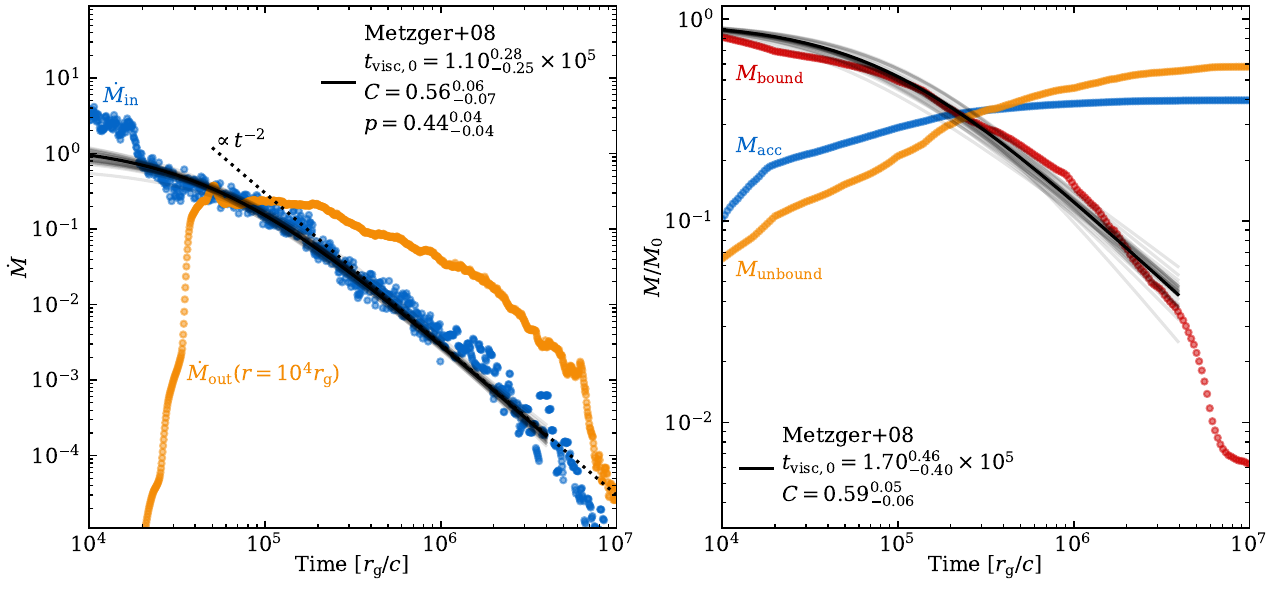}
\caption{Smoothed long-term evolution of an FM torus initially between $20\,r_\mathrm{g}$ and $100\,r_\mathrm{g}$ around a black hole with $a=0$. Left: history of accretion rate around the horizon and outflow rate. The black line is the best-fit model of accretion rate in \citet{Metzger2008MNRAS.390..781M} and the gray lines are the scatter. The accretion rate decreases with time following $\dot{M}\propto t^{-2}$ when $t\gg t_\mathrm{visc,0}$. Right: Bound mass, accreted mass, and unbound mass as a function of time. 
\label{fig:secular}}
\end{figure*}

\section{Validation} \label{sec:validation}

In this section, we present a set of validation simulations. First, we perform tests on problems with known analytic solutions including spherically symmetric Bondi accretion and black hole energy extraction with monopole magnetic field. Then we test problems without known analytic solutions, including magnetized Bondi accretion, torus accretion, and long-term secular evolution of tori. These problems are ideal testbeds for our method. For each case, we compare two runs: a ``standard'' run, i.e., a standard GRMHD simulation, and a ``cyclic zoom'' run using the cyclic zoom method we described above. All runs in this section use a sufficiently large cubic box of size $[-2^{17},2^{17}]^3\approx[-1.3\times10^5,1.3\times10^5]^3\,r_\mathrm{g}^3$ with a root grid of $128^3$ cells and 15 levels of mesh refinement, with the finest resolution being $\Delta x_\mathrm{min}=1/16\,r_\mathrm{g}$ except for the monopole test, which only covers the domain of $z>0$. The main parameters of these runs are summarized in \tab\ref{tab:runs}. Below, we discuss the details of each test.

\subsection{Hydrodynamic Bondi Accretion} \label{subsec:valid_hyd_bondi}
Spherically symmetric, adiabatic Bondi accretion onto a Schwarzschild black hole~\citep{Michel1972Ap&SS..15..153M} is a common test problem for GR hydrodynamics codes. Similarly to \citet{Stone2024arXiv240916053S}, we set the initial conditions using the solution in~\citet{Hawley1984ApJ...277..296H}. We use an adiabatic index of $\gamma_\mathrm{ad}=5/3$, an adiabat $K=p/\rho^{\gamma_\mathrm{ad}}=1$, and a sonic point radius $r_\mathrm{c}=16\, r_\mathrm{g}$~\citep[the same as][]{Cho2024ApJ...977..200C}, corresponding to a Bondi radius $r_\mathrm{B}=2GM/c_{s,\infty}^2=2GM/(\gamma_\mathrm{ad} T_\infty)\approx800\, r_\mathrm{g}$\footnote{Some literature defines the Bondi radius to be $r_\mathrm{B}=GM/c_{s,\infty}^2$, i.e, half of the Bondi radius defined here.} and thus a Bondi time $t_\mathrm{B}=r_\mathrm{B}/c_{s,\infty}\approx1.5\times10^4t_\mathrm{g}$. We use a cubic box of size $[-2^{17},2^{17}]^3\approx[-1.3\times10^5,1.3\times10^5]^3\,r_\mathrm{g}^3$ with a root grid of $128^3$ cells and 15 levels of mesh refinement with a finest resolution of $\Delta x_\mathrm{min}=1/16\,r_\mathrm{g}$. The outer boundary conditions are prescribed through the analytic solution. The inner boundary is the excision region detailed in \citet{Stone2020ApJS..249....4S}. For the cyclic zoom, we set $n=8$ and 
\begin{equation}
    t_{\mathrm{run},i}=
    \begin{cases}
      100 \,t_\mathrm{g} & \text{if $i=0$},\\
      0 & \text{if $i=1,2$},\\
      10\, t_{\mathrm{z},i}& \text{otherwise},
    \end{cases}
\end{equation}
where the local dynamical time $t_{\mathrm{z},i}=r_{\mathrm{z},i}/\sqrt{GM/r_{\mathrm{z},i}}$. We run the simulation to $10^5\,t_\mathrm{g}\approx 6\,t_\mathrm{B}$, and its main parameters are summarized in \tab\ref{tab:runs}.

\figu\ref{fig:hyd_bondi} (left panel) shows the $z=0$ density slice at the end of the simulation. Despite the Cartesian grid and multiple times of mesh refinements and derefinements, the solution maintains nearly perfect symmetry. The radial profile of the accretion rate averaged in time over the later half of the simulation is shown in the right panel of~\figu\ref{fig:hyd_bondi}. The accretion rate is nearly constant over the whole simulation domain. It is consistent with the predicted Bondi accretion rate within $\lesssim 0.1\%$. This proves that the cyclic zoom method can recover the analytic solution with high precision. 

\subsection{Blandford-Znajek Monopole} \label{subsec:valid_monopole}
The \citet{Blandford1977MNRAS.179..433B} (BZ) monopole test problem is another common test problem for GRMHD codes. Following~\citet{Chael2024MNRAS.532.3198C} and \citet{Stone2024arXiv240916053S}, a monopole magnetic field $B^r\propto1/r^2$ (initialized using a toroidal vector potential) threads a spherically symmetric medium about a Kerr black hole with spin $a=0.5$. A density power law $\rho(r)=B^2/\sigma_c\propto r^{-4}$ with $\sigma_c=10$ is prescribed for the medium. The initial pressure is set to $p=\rho T$ where $T=\gamma_\mathrm{ad}-1$ with an adiabatic index $\gamma_\mathrm{ad}=4/3$. A ceiling of magnetization $\sigma_\mathrm{max}=10$ is applied for stability. We use a box of $x/r_\mathrm{g}\in[-2^{14},2^{14}]$, $y/r_\mathrm{g}\in[-2^{14},2^{14}]$, $z/r_\mathrm{g}\in[0,2^{14}]$ with a root grid of $128\times128\times64$ cells and 12 levels of mesh refinement with a finest resolution of $\Delta x_\mathrm{min}=1/16\,r_\mathrm{g}$. Outflow boundary conditions are applied everywhere except at the $z=0$ plane, where we instead impose reflecting boundary conditions. For the cyclic zoom run, we set $n=6$ and 
\begin{equation}
    t_{\mathrm{run},i}=
    \begin{cases}
      100 \, t_\mathrm{g} & \text{if $i=0$},\\
      0 & \text{if $i=1,2,3$},\\
      10 \, t_{\mathrm{z},i}& \text{otherwise},
    \end{cases}
\end{equation}
where $t_{\mathrm{z},i}=r_{\mathrm{z},i}/c$. We evolve the system for $4\times10^3\,t_\mathrm{g}$. The main parameters are listed in \tab\ref{tab:runs}. 

The monopole test challenges the algorithm because the dynamics is governed nearly entirely by the magnetic field. Thus, it is a particularly good test of our method for evolving the magnetic field in the mask region. One key diagnostic is the field rotation rate,
\begin{equation}
    \Omega_\mathrm{F}=\frac{b^ru^\phi-b^\phi u^r}{b^r u^t - b^t u^r}
\end{equation}
after mesh derefinement and refinement. \figu\ref{fig:mono_valid} plots the $z=0$ slice for both the standard and cyclic zoom run. Visual inspection shows that the rotational symmetry is preserved even after $3000\,t_\mathrm{g}$. More quantitatively, we plot the history of the field rotation rate relative to the horizon rotation rate $\Omega_\mathrm{H}$ and its time-averaged radial profile. The rotation rate is consistent with the expected solution $\Omega_\mathrm{F}/\Omega_\mathrm{H}=1/2$.

In addition, we plot the history and radial profiles of the magnetic flux and the energy flux in \figu\ref{fig:mono_valid}. 
The magnetic flux slowly decays due to diffusion with a characteristic timescale of $\sim O(10^3\,t_\mathrm{g})$. We find one can improve flux conservation for this test by enforcing the EMF $\mathcal{E}_x=\mathcal{E}_y=0$ at the $z=0$ plane. However, we would never be able to use such a fix in any practical applications, since it relies on symmetry across the equatorial plane. Even for this monopole test, the initial condition is not perfectly symmetric, as there is a net flux of $B_z$ crossing the $z=0$ plane in the center within the horizon, since we initialize the magnetic field using vector potential, which prohibits a true monopole inside the simulation domain. The energy feedback efficiency correspondingly decays with time. For the cyclic zoom run, the flux decays slightly more, likely because of the extra diffusion. We confirm this by running a standard run with half of the resolution, which shows decays of magnetic flux and energy flux similar to those of the cyclic zoom run. Therefore the extra decay is induced by the relatively poor resolution, not the cyclic zoom method itself. Still, the cyclic zoom run is consistent with the standard run within a factor of $\sim 2$ over thousands of dynamical times (the light-crossing times) in the inner region. The issue of magnetic field decay will be alleviated in the accretion problem, where plasma inflow helps to hold the magnetic flux at the horizon, as we shall see in the following tests.

\subsection{Magnetized Bondi Accretion} \label{subsec:valid_bondi}
We here model magnetized Bondi accretion onto black holes following a series of previous studies~\citep{Lalakos2022ApJ...936L...5L, Cho2023ApJ...959L..22C, Cho2024ApJ...977..200C, Galishnikova2025ApJ...978..148G}. This problem has no known analytical solution and requires numerical simulations. In the presence of strong magnetic fields, the accretion flow becomes turbulent. Therefore, it is a good test problem for our technique.

As an initial condition, instead of the analytic Bondi solution, we specify a uniform background with constant density $\rho_\infty$ and temperature $T_\infty$, and a constant vertical magnetic field along the $z$-direction with constant plasma $\beta\equiv p_\mathrm{gas}/p_\mathrm{mag}$, where $p_\mathrm{mag}=b^2/2$. In these runs, following \citet{Galishnikova2025ApJ...978..148G}, we use an adiabatic index $\gamma_\mathrm{ad}=5/3$ and a Bondi radius $r_\mathrm{B}=500\, r_\mathrm{g}$, and thus a Bondi time $t_\mathrm{B}\approx8\times10^3t_\mathrm{g}$. We test spins $a=0$  and $0.9375$, for $\beta = 1$ and $1000$. The internal energy is perturbed by white noise with a maximum amplitude of $2\%$ to break the symmetry. There are thus four pairs of runs in total, with their main parameters summarized in \tab\ref{tab:runs}.

In the cyclic zoom runs, we set $n=8$ and 
\begin{equation}
    t_{\mathrm{run},i}=
    \begin{cases}
      500 \,t_\mathrm{g} & \text{if $i=0$},\\
      0 & \text{if $i=1,2$},\\
      \min\left(50\, t_{\mathrm{z},i}, 0.15\,t_\mathrm{B}\right) & \text{otherwise}.
    \end{cases}
\end{equation}
The runtime at level $0$ is $t_{\mathrm{run},i}=500\,t_\mathrm{g}$ so that the turbulent fluid and the magnetic field have enough time to relax to a quasi-steady state. The runtime is $50$ times the local dynamical time $t_{\mathrm{z},i}=r_{\mathrm{z},i}/\sqrt{GM/r_{\mathrm{z},i}}$ so that the turbulent flow can communicate the information sufficiently. Finally, we cap the runtime at $0.15\,t_\mathrm{B}$ to capture the variability within the Bondi timescale. In this way, we reach a speedup of $\gtrsim 10$. All test runs are evolved for $2\times10^5\, t_\mathrm{g}\approx25\,t_\mathrm{B}$.

The results for the cases with $a=0$ and $\beta=1$ are shown in \figu\ref{fig:bondi_valid}. The accretion flow is turbulent and establishes a quasi-steady state with accretion rate $\dot{M}\sim 0.2 \dot{M}_\mathrm{B}$, energy feedback rate $\dot{E}\sim 2\times10^{-3} \dot{M}_\mathrm{B}c^2$ with efficiency $\eta\approx1\%$, and saturated magnetic flux of $\phi_\mathrm{BH}\sim 40$. The upper panel of \figu\ref{fig:bondi_valid} plots the $y=0$ slices of plasma-$\beta$ from small to large scales. The cyclic zoom run reproduces qualitatively similar accretion flows and magnetized outflows. Since the fluid is dynamic and turbulent, we do not expect the system to be in the same state, but we look for statistically similar behaviors. In fact, even the standard run can show different snapshots after such a long duration, if we rerun it. From the statistical point of view, the cyclic zoom method captures the mass accretion rate, the energy feedback rate, and the magnetic flux parameter very well. As the accretion flow is dynamic and turbulent, we plot the spherically and time-averaged radial profiles of the accretion flow in \figu\ref{fig:bondi_valid}. There is excellent consistency between the standard run and the cyclic zoom run. As indicated by the accretion rate, the accretion flow builds a steady state within the Bondi radius through the cyclic zoom method, but the corresponding simulation only costs less than $10\%$ of the computational resources. The variability of the quantities is also similar between the two runs.

The upper panel of \figu\ref{fig:bondi_valids} plots $y=0$ slices of plasma-$\beta$ for all four pairs of runs at nearly the end of the simulations. All the runs show qualitatively similar accretion flows. The lower panel of \figu\ref{fig:bondi_valids} shows the time-averaged accretion rate $\dot{M}$, the magnetic flux parameter $\phi_\mathrm{BH}$, and the feedback efficiency $\eta$, as well as their distributions in the four pairs of runs. Overall, the cyclic zoom runs produce very similar results to the standard runs. Note that, for $a=0$ and $\beta=10^3$, the magnetic field is too weak to produce positive feedback, whereas the case with $a=0.9375$ and $\beta=10^3$ has a relatively larger variability, as found by \citet{Galishnikova2025ApJ...978..148G}. Thus, this run is more challenging to reproduce accurately, yet the cyclic zoom method does well.

\subsection{Torus Accretion} \label{subsec:valid_torus}
The accretion of a torus onto a black hole is another standard test problem for GRMHD codes~\citep[e.g.,][]{Porth2019ApJS..243...26P}. Following \citet{Narayan2022MNRAS.511.3795N}, we perform a suite of simulations of magnetically arrested disks (MAD) with different spins, $a=-0.9375$, $-0.5$, $0$, $0.5$, and $0.9375$. We initialize the simulations with the \citet{Fishbone1976ApJ...207..962F} torus solution, which is determined by four parameters: the inner edge of the torus $r_\mathrm{in}$, the location of the pressure maximum $r_\mathrm{peak}$, the adiabatic index of the fluid $\gamma_\mathrm{ad}$, and the maximum density $\rho_\mathrm{max}$. The location of the outer edge of the torus is a function of these parameters. We set $r_\mathrm{in}=20\,r_\mathrm{g}$ for all cases and adjust $r_\mathrm{peak}$ according to $a$ so that $r_\mathrm{out}\approx10^4\,r_\mathrm{g}$. We adopt an ideal gas equation of state with an adiabatic index of $\gamma_\mathrm{ad}=13/9$.

The magnetic field is initialized with a weak single large magnetic field loop defined by the poloidal vector potential $A_\phi$:
\begin{equation}
\begin{aligned}
    A_\phi &= \max(0,q),\\ 
    q &= \left[\left(\frac{\rho}{\rho_\mathrm{max}}\right)\left(\frac{r}{r_\mathrm{in}}\right)^3e^{-r/r_\mathrm{mag}}\sin^3\theta\right]-A_{\phi,\mathrm{cut}},
    \label{eq:torus_Aphi}
\end{aligned}
\end{equation}
where we set $r_\mathrm{mag}=400\,r_\mathrm{g}$ and $A_{\phi,\mathrm{cut}}=0.2$. After determining the field components via the vector potential of \eq\ref{eq:torus_Aphi}, we normalize the initial magnetic field strength in the disk following \citet{Porth2019ApJS..243...26P} such that the maximum gas pressure and the maximum magnetic pressure in the torus (which do not necessarily occur at the same location) satisfy $\beta_\mathrm{max}\equiv(p_\mathrm{gas})_\mathrm{max}/(p_\mathrm{mag})_\mathrm{max}=100$. The internal energy is perturbed by white noise with a maximum amplitude of $2\%$ to break symmetry and excite the magnetorotational instability (MRI) inside the torus.

For the cyclic zoom runs, we set $n=6$ and the runtime
\begin{equation}
    t_{\mathrm{run},i}=
    \begin{cases}
      500\,t_\mathrm{g} & \text{if $i=0$},\\
      0 & \text{if $i=1,2$},\\
      100\, t_{\mathrm{z},i} & \text{otherwise},
    \end{cases}
\end{equation}
where $t_{\mathrm{z},i}=r_{\mathrm{z},i}/c$ since the fastest speed on all scales is the speed of light due to the presence of relativistic jets. Each simulation is evolved for $10^5\,t_\mathrm{g}$ to reach a statistically steady state. In this way, we reach a speedup of $\sim 10$. The details of the parameters and runtime are summarized in \tab\ref{tab:runs}.

\figu\ref{fig:torus_valid} compares the cyclic zoom run with the standard run for the fast-spinning case with $a=0.9375$. The cyclic zoom run reproduces similar accretion flows, jets, and all the key quantities. It reaches a flat mass flux within $\sim 200\, r_\mathrm{g}$, similarly to the standard runs. The mass outflow profile is also very similar to the standard results, indicating that the large-scale evolution of the torus is also correctly captured. The accretion flow reproduces the strong BZ jet with a similar feedback efficiency of $\sim 100\%$, as well as similar dispersion ranging from $\sim 60\%$ to $\sim 200\%$. The feedback efficiency profiles are flat within $\sim 10^5r_\mathrm{g}$, similarly to those in the standard run. The system is in a MAD state with the normalized magnetic flux parameter around the horizon $\phi_\mathrm{BH}\sim 50$, similarly to that in the standard run.

Time-averaged radial profiles of the accretion rate, magnetic flux parameter, feedback efficiency, and specific angular momentum flux for different spins are shown in \figu\ref{fig:torus_valids}. The cyclic zoom runs reach a steady state within $\sim 200\, r_\mathrm{g}$ for all spins. The dependence of these variables on spins is correctly captured within $\sim 10\%$ for most cases. The one that is not so satisfying is the case with $a=0$, where the efficiency is overestimated by a factor of $\sim 2$. This is because the Poynting flux is slightly overestimated in the cyclic zoom method. But still, the feedback power is much better here than in previous works using analytic models, as they cannot obtain the accretion rate accurately~\citep[e.g.,][]{Weinberger2025arXiv250213241W}.

To demonstrate the dependence of the tests on the BH spin, we further show the distribution of accretion rate, magnetic flux parameter, and feedback efficiency as a function of spin in \figu\ref{fig:torus_valids}. The accretion rate has a weak dependence on the BH spin. The magnetic flux parameter follows the functional fit from \citet{Narayan2022MNRAS.511.3795N} \citep[see also][]{Tchekhovskoy2011MNRAS.418L..79T, Chael2025MNRAS.537.2496C} based on a series of torus accretion simulations. The feedback efficiency also follows the formula fitted by \citet{Narayan2022MNRAS.511.3795N} plus a constant efficiency of $3\%$ as an estimate of the feedback from the accretion disk itself. Overall, the cyclic zoom method can reproduce the dependence of these quantities on spin.

To highlight that the inclusion of $\delta \boldsymbol{\mathcal{E}}$ is important for preserving the outgoing Poynting flux in the case of rotating black holes, we perform two extra cyclic zoom runs with and without the source term $\delta \boldsymbol{\mathcal{E}}$ in \eq\ref{eq:induction_delta_emf}. The cyclic zoom runs are restarted from the standard run of the FM torus around a spinning black hole with $a=0.9375$ at a time when it expands to larger radii and stays in a quasi-steady state ($t=5\times10^4\,t_\mathrm{g}$). For the cyclic zoom, we set $n=8$ and the runtime
\begin{equation}
    t_{\mathrm{run},i}=
    \begin{cases}
      500\,t_\mathrm{g} & \text{if $i=0$},\\
      0 & \text{if $i=1,2$},\\
      100\, t_{\mathrm{z},i} & \text{otherwise},
    \end{cases}
\end{equation}
where $t_{\mathrm{z},i}=r_{\mathrm{z},i}/c$. The history of the feedback power in \figu\ref{fig:torus_valid_emf} shows that, without $\delta\boldsymbol{\mathcal{E}}$, the Poynting flux is reduced by $\sim 30\%$ each time we zoom out if the horizon is not resolved. It is $\sim 10$ times smaller than the standard run when we zoom out to large scales ($r_{\mathrm{z},i}=128\,r_\mathrm{g}$). The decline of the Poynting flux leads to a smaller feedback power on large scales. On the other hand, the cyclic zoom run that includes the source term from small scales can maintain the Poynting flux even if $r_{\mathrm{z},i}$ is large.

\subsection{Long-term Evolution of Torus Accretion} \label{subsec:valid_secular}
Apart from the statistically steady states, we also aim to simulate the long-term evolution of accreting systems using the cyclic zoom method, which is computationally not feasible in current GRMHD simulations due to the very long viscous timescale compared with the dynamical timescale, i.e., $t_\mathrm{visc}\gtrsim 10^3 \, t_\mathrm{dyn}$. We use a setup similar to \sect\ref{subsec:valid_torus} with spin $a=0$ or $0.9375$, and $r_\mathrm{in}=20\,r_\mathrm{g}$, except that we use a smaller outer radius $r_\mathrm{out}=100\,r_\mathrm{g}$ so that the torus mass is concentrated around $\sim 50\, r_\mathrm{g}$. Such a radially concentrated torus produces significant secular evolution on resolvable timescales. The cyclic zoom parameters are the same as those of the tori studied in~\sect\ref{subsec:valid_torus}.

To study the secular evolution, we run the standard simulations for a very long time of $10^6\, t_\mathrm{g}$ for the case with $a=0$, and for $2\times 10^5\, t_\mathrm{g}$ for the case with $a=0.9375$. The history of the accretion rate, the feedback power, and the magnetic flux parameter are shown in \figu\ref{fig:secular_valid}. The cyclic zoom method can capture the secular evolution of the torus very well over a factor of $\sim 100$ in accretion rate for $a=0$. Therefore, apart from steady or statistically steady states, the cyclic zoom method is also suitable for systems undergoing secular evolution. We will further analyze the longer-term dependence and discuss the details of the secular evolution in~\sect\ref{subsec:app_secular}.

\section{Applications} \label{sec:results}

In this section, we present our results for a series of problems that are computationally not feasible in the foreseeable future using standard GRMHD simulations. In particular, we consider Bondi accretion with a large Bondi radius of $4\times10^5\,r_\mathrm{g}$, accretion of a giant torus, and very long-term evolution of a torus.

\subsection{Magnetized Bondi Accretion over Galactic Scales} \label{subsec:app_bondi}

Here, we simulate the accretion of magnetized plasma with a Bondi radius $r_\mathrm{B}=4\times10^5\, r_\mathrm{g}$, $\beta = 1$ and spin $a=0$ and $a=0.9375$ for about one billion $t_\mathrm{g}$ ($\approx 5 \,t_\mathrm{B}$ where $t_\mathrm{B}\approx1.8\times10^8\, t_\mathrm{g}$). The corresponding temperature is $T_\infty\approx 2\times10^7\,\mathrm{K}$, similar to the typical temperature of the hot plasma in elliptical galaxies~\citep{Russell2015MNRAS.451..588R}. The simulation time translates to $\sim 1\,\mathrm{Myr}$ for M87*-like SMBHs~\citep{M87EHT_I_2019ApJ...875L...1E}. We use a cubic box of size $[-2^{24},2^{24}]^3\approx[-1.7\times10^7,1.7\times10^7]^3\,r_\mathrm{g}^3$ with a root grid of $128^3$ cells and 22 levels of mesh refinement. For the cyclic zoom, we set $n=15$ and use the runtime
\begin{equation}
    t_{\mathrm{run},i}=
    \begin{cases}
      500\,t_\mathrm{g} & \text{if $i=0$},\\
      0 & \text{if $i=1,2$},\\
      \min\left(20\, t_{\mathrm{z},i},\, 0.15\,t_\mathrm{B}\right) & \text{otherwise},
    \end{cases}
\end{equation}
where the local dynamical time is $t_{\mathrm{z},i}=r_{\mathrm{z},i}/\sqrt{GM/r_{\mathrm{z},i}}$.

\figu\ref{fig:bondi_slice} shows slices across the $y=0$ plane for density and plasma-$\beta$, for the cases with $a=0$ and $a=0.9375$ at a later stage. The symmetry of Bondi accretion is clearly broken on all scales within the Bondi radius. We note that the flow structure varies from time to time for both cases, with jets appearing and disappearing. Overall, there is typically stronger feedback for the spinning case. The jets are often ``choked'' on the mesoscales interior to the Bondi radius. The feedback does not show a preference for a certain direction within the Bondi radius, even for a highly spinning black hole. This indicates that it is hard for the relativistic jets to propagate out to the Bondi radius. Fluid rotation may be required for a persistent jet~\citep{Galishnikova2025ApJ...978..148G}.

\figu\ref{fig:bondi_application} shows the evolution and time-averaged radial profiles of key variables including the accretion rate, the energy feedback rate, and the magnetic flux parameter. The accretion rate is $\sim 1\%\, \dot{M}_\mathrm{B}$ for $a=0$ and $\sim 0.5\%\dot{M}_\mathrm{B}$ for $a=0.9375$. The net accretion rate is constant within $\sim 10^3\,r_\mathrm{g}$ but slightly increases going to larger radii. This is partially because of the relatively short time of evolution for the average. In addition, it is very hard to find a flat accretion profile, as both the inflow and the outflow are very strong. As a result, the accretion rate exhibits considerable variability, which is higher when the black hole is spinning. The feedback efficiency is $\sim 1\%$ for $a=0$ and $\sim 10\%$ for $a=0.9375$. We note that, for $a=0.9375$, the feedback has a noticeably stronger variability with feedback efficiency ranging from $\sim 1\%$ to $\sim 30\%$. For the feedback, we also see a gradual transition from EM energy flux to hydrodynamic energy flux at larger radii. Within the Bondi radius, the mass inflow rate $\dot{M}_\mathrm{in}\propto r^{1/2}$ and the density profile follows $\rho\propto r^{-1}$, consistent with the suppression of the net accretion rate. This radial dependence of the density may be universal, as is reported in various contexts and with varying physics~\citep{Ressler2018MNRAS.478.3544R, Ressler2020ApJ...896L...6R, Xu2019MNRAS.488.5162X, Xu2023ApJ...954..180X, Guo2020ApJ...901...39G, Guo2023ApJ...946...26G, Guo2024ApJ...973..141G, Lalakos2022ApJ...936L...5L, Cho2023ApJ...959L..22C, Cho2024ApJ...977..200C}. The gas remains moderately magnetized, with an azimuthally averaged plasma-$\beta\sim1$ at all radii.

The accretion rates and feedback efficiencies for accretion with different Bondi radii found in this work are summarized in \figu\ref{fig:bondi_scaling}， including two extra runs with an intermediate Bondi radius of $r_\mathrm{B}=2\times10^4\,r_\mathrm{g}$ (see \tab\ref{tab:runs}). Overall, we find a good agreement with a scaling according to $\dot{M}_\mathrm{acc}/\dot{M}_\mathrm{B}\approx(10r_\mathrm{g}/r_\mathrm{B})^{1/2}$. The feedback efficiency is roughly $\sim 1\%$ for the Schwarzschild black hole and $\sim 10\%$ for the highly spinning black hole. Note that, for the spinning black hole, the feedback efficiency has a larger dispersion, reflecting the higher variability in this case, consistent with~\citet{Galishnikova2025ApJ...978..148G}. We note that a more frequent sampling on small scales may help to better capture this variability.

\subsection{Accretion of Torus over Galactic Scales} \label{subsec:app_torus}
Here, we evolve a more realistic torus initially located between $r_\mathrm{in}=10^3\,r_\mathrm{g}$ and $r_\mathrm{out}=10^4\,r_\mathrm{g}$ around a black hole with spin $a=0.9375$. Correspondingly, we set a larger exponential falloff radius for the vector potential of $r_\mathrm{mag}=4000\,r_\mathrm{g}$. We use a cubic box of size $[-2^{24},2^{24}]^3\approx[-1.7\times10^7,1.7\times10^7]^3\,r_\mathrm{g}^3$ with a root grid of $128^3$ cells and 22 levels of mesh refinement. For a cyclic zoom, we set $n=11$ and use a runtime 
\begin{equation}
    t_{\mathrm{run},i}=
    \begin{cases}
      500\,t_\mathrm{g} & \text{if $i=0$},\\
      0 & \text{if $i=1,2$},\\
      100\, t_{\mathrm{z},i} & \text{otherwise},
    \end{cases}
\end{equation}
where $t_{\mathrm{z},i}=r_{\mathrm{z},i}/c$. We evolve the torus for 10 million $t_\mathrm{g}$ to reach a statistically steady state.

The evolution of key variables and time-averaged radial profiles of the accretion flow is shown in \figu\ref{fig:torus_application}. We reach a quasi-steady state within $\sim 2\times10^3\,r_\mathrm{g}$, as is shown in the radial profile of the accretion rate. Due to the presence of both strong inflow and outflow of the torus, it is hard to get a very flat profile. In addition, the accretion rate may not be necessarily constant, since the torus is gradually evolving. The feedback efficiency profile is flat within $\sim 10^7\,r_\mathrm{g}$ with a mean value of $50\%$ fluctuation between $\sim 10\%-100\%$. The density profile follows $\rho\propto r^{-1}$ within $10^3\,r_\mathrm{g}$. The magnetic flux parameter is still $\phi_\mathrm{BH}\sim 50$, meaning that the accretion flow is still a MAD. This test challenges the cyclic zoom method's ability to maintain the jet even if the horizon is completely unresolved as the finest resolution becomes $\Delta x=64\,r_\mathrm{g}$ when $i=10$. It also confirms that a giant torus located at $\sim10^3\,r_\mathrm{g}$ or farther around a spinning black hole is able to produce a strong, sustained jet over a long timescale.

\subsection{Long-term Evolution of Torus Accretion} \label{subsec:app_secular}

In this subsection, we track the evolution of a torus until almost all of the mass is either accreted onto the black hole or turned into outflows capable of escaping to infinity. We continue running the torus presented in \sect\ref{subsec:valid_secular} with $r_\mathrm{in}=20\,r_\mathrm{g}$, $r_\mathrm{out}=100\,r_\mathrm{g}$ and spin $a=0$. After $10^6 \,t_\mathrm{g}$, as the torus expands to a larger scale, we increase the number of levels from $n=6$ to $n=8$ to further speed up the simulation. Then, after $3\times10^6\, t_\mathrm{g}$, we again increase $n$ to $n=10$ and evolve it to $10^7\, t_\mathrm{g}$.

\figu\ref{fig:secular} shows the long-term evolution of the torus accretion, including the evolution of mass and accretion rate. We separate the mass into bound ($\mathcal{B}_\mathrm{gas}<0$) and unbound ($\mathcal{B}_\mathrm{gas}>0$) using the hydrodynamic Bernoulli parameter:
\begin{equation}
    \mathcal{B}_\mathrm{gas}=-u_t(\rho+u+p)/\rho-1.
\end{equation}
As shown in \figu\ref{fig:secular}, $\sim 40\%$ of the mass of the torus is accreted, while the rest is turned into outflow that escapes to infinity. The accretion rate gradually decreases from $\sim1$ to $\sim 10^{-4}$.

The secular evolution of an accretion flow with fixed scale height $h/r$, conserved total angular momentum, and no outflows follows $\dot{M}\propto t^{-4/3}$. \citet{Metzger2008MNRAS.390..781M} provided a series of solutions considering mass and angular momentum loss. The outflow is parameterized using
\begin{equation}
    \dot{M}_\mathrm{out}=\left[1-\left(\frac{r_*}{r_\mathrm{d}}\right)^p\right]\frac{fM_\mathrm{d}}{t_\mathrm{visc}},
\end{equation}
where $r_*$ is the radius of the no-torque boundary condition, $r_\mathrm{d}$ is the radius of the disk, $M_\mathrm{d}$ is the disk mass, and $t_\mathrm{visc}$ is the viscous time. The angular momentum loss rate from the disk is
\begin{equation}
    \dot{J}=-C\dot{M}_\mathrm{out}(GMr_\mathrm{d})^{1/2},
\end{equation}
where $C$ is a constant that depends on the torque exerted by the outflowing mass on the remaining disk. For $C(r_*/r_\mathrm{d})^p\ll 1-C$, the disk mass evolves as
\begin{equation}
    M_\mathrm{d}\simeq M_\mathrm{d,0}\left[1+3f(1-C)\left(\frac{t}{t_\mathrm{visc,0}}\right)\right]^{-1/[3(1-C)]},
\end{equation}
and the accretion rate follows
\begin{equation}
\begin{aligned}
    \dot{M}_\mathrm{acc} \simeq & f\frac{M_\mathrm{d,0}}{t_\mathrm{visc,0}}\left(\frac{r_*}{r_\mathrm{d,0}}\right)^p\\
    & \times\left[1+3f(1-C)\left(\frac{t}{t_\mathrm{visc,0}}\right)\right]^{-\frac{1}{3(1-C)}-1-\frac{2p}{3}}.
\end{aligned}
\end{equation}
We use $f=1.6$ following \citet{Metzger2008MNRAS.390..781M}, and fix $r_*=6\,r_\mathrm{g}$ and $r_\mathrm{d,0}=50 \,r_\mathrm{g}$.

Using the formula in \citet{Metzger2008MNRAS.390..781M}, we fit the evolution of disk mass and accretion rate respectively. We only fit for $10^4<t/t_\mathrm{g}<4\times10^6$ since the MRI is not fully developed at early times and the gas is too loosely bounded to be thought of as a disk at late times. For evolution of the accretion rate, we find $t_\mathrm{visc,0}=1.10_{-0.25}^{0.28}\times 10^5\,t_\mathrm{g}$, $C=0.56_{-0.07}^{0.06}$, and $p=0.44_{-0.04}^{0.04}$. For evolution of disk mass, we use the bound mass as a proxy of the disk mass and find $t_\mathrm{visc,0}=1.70_{-0.40}^{0.46}\times 10^5\,t_\mathrm{g}$ and $C=0.59_{-0.06}^{0.05}$. The fits are overplotted in \figu\ref{fig:secular}. These two fits are consistent with each other, implying a result of $C\approx1/2$ and $p\approx1/2$. Note also that $t_{\rm visc,0} \sim 10^{5}\,t_{\rm g}$ corresponds to $t_{\rm visc,0} \sim 300 \, r_{\rm d,0}^{3/2}$ which is consistent with $h/r \sim 1/3$ and a dimensionless stress $\alpha \sim 0.03$. The fits found here correspond to a late-time evolution of the bound disk mass and accretion rate scaling as $M_\mathrm{disk}\propto t^{-2/3}$ and $\dot{M}_\mathrm{in}\propto t^{-2}$. We will analyze additional details of this evolution in future work.

\section{Discussion}\label{sec:discussion}
The tests presented above demonstrate that the cyclic zoom method is able to connect large scales and small scales for various problems and different initial conditions related to gas flows around compact objects. It is able to pass the information from small scales to large scales and vice versa, preserving predictions for mass and energy fluxes in accretion problems. Below, we discuss the limitations, advantages, comparisons to previous work, and future applications of this technique.

\subsection{Difficulties and Limitations}\label{subsec:diff}
Given the turbulent nature of the accretion flow, there are intrinsic difficulties in comparing standard runs and the cyclic zoom runs at the ``same'' time. Even for standard simulations, the details of a numerical solution can be different if we compute it again when the fluid is turbulent, because the chaotic nonlinear dynamics can amplify tiny differences originating in small variations in floating-point roundoff. That is why we compare statistical properties, such as averaged values and their dispersion, in most cases. 

Our method effectively ``freezes'' the small-scale physics to control the accretion onto small scales and power the feedback to larger scales. This means that the short-time variability of the real physical system may not be sufficiently captured. More frequent sampling, or different strategies for carrying out the cyclic zoom, like ``F'' or ``W'' patterns in the spacetime domain, may help to capture more of the small-scale variability. In addition, the evolution of the mask region can be made more sophisticated to capture more of the small-scale information. For example, instead of using the variables right before the mesh refinement, one may use the time-averaged variables or a more data-driven model. There might also be some room for investigating the optimal size of the mask region in higher-resolution simulations.

\subsection{Speedup} \label{subsec:speedup}
\figu\ref{fig:method} illustrates the speedup that can be obtained with the cyclic zoom method. For a normal GRMHD simulation using AMR or spherical coordinates, the computation time, $t_\mathrm{comp}$, scales with the characteristic length scale of interest, $r_\mathrm{char}$, as $t_\mathrm{comp}\propto r_\mathrm{char}^{3/2}$. This is because the characteristic time varies as $t_\mathrm{char}\propto r_\mathrm{char}/v_\mathrm{char} \propto r_\mathrm{char}^{3/2}$, where the characteristic speed scales as $v_\mathrm{char}\propto r_\mathrm{char}^{-1/2}$. Using the cyclic zoom method, we can ideally achieve $t_\mathrm{comp}\propto\log r_\mathrm{char}$ and thus a speedup of $\propto r_\mathrm{char}^{3/2}$ since the simulation spends a similar amount of time on all scales in terms of characteristic timescale. 

However, in the presence of material with relativistic speed on large scales, e.g., jets due to a spinning black hole, we have $t_\mathrm{comp}\propto r_\mathrm{char}^{1/2}$ using cyclic zoom and thus a speedup $\propto r_\mathrm{char}$. Therefore, the actual computation time can typically be expected to lie between these two limits, i.e., $t_\mathrm{comp}\propto r_\mathrm{char}^{\sim 0-0.5}$, and thus a realistic speedup is $\propto r_\mathrm{char}^{\sim 1-1.5}$, as shown by the cyclic zoom runs in \figu\ref{fig:method}. In addition, since we typically spend more time if $i=0$, the scaling of the cyclic zoom method has a relatively large coefficient in front, due to the corresponding overhead. Depending on how many levels we are using for the cyclic zoom, and how much time we spend on each level, the speedup can be higher or lower by factors of a few. Given the computational resources available, we can flexibly choose how much high-frequency information we want to obtain, as long as we keep $r_{\mathrm{z},\,\max}\lesssim r_\mathrm{char}$ and $t_{\mathrm{run},\,\max}\lesssim t_\mathrm{char}$. Overall, we can achieve a speedup of $\gtrsim 10^5$ for galactic-scale simulations. Note that this allows simulations that would otherwise take more than hundreds of years to run to be effectively completed within a few days or weeks.

\subsection{Comparison to previous work}
The work presented here was inspired by \citet{Cho2023ApJ...959L..22C}, who presented a ``multi-zone'' method for bridging scales and tested it for Bondi accretion. They successfully spanned 7 orders of magnitude in radius and simulated accretion onto a non-spinning SMBH from an external medium with a Bondi radius $\gtrsim 10^5\, r_\mathrm{g}$. \citet{Cho2024ApJ...977..200C} further applied the method to simulations using initial conditions from a large-scale galaxy simulation and achieved a steady state over 8 decades in radius. Our method preserves the spirit of ``V-cycle'' or ``$\Lambda$-cycle'' as in their method.

However, we treat the magnetic field in a different manner. We still evolve the magnetic field $\boldsymbol{B}$ in the mask region, while \citet{Cho2024ApJ...977..200C} hold the magnetic field fixed when the zone is not active. In this way, our method can avoid the strong shear of the magnetic field around the boundary when there is rotation and inhomogeneity. The shear of the magnetic field may not be severe when the flow is uniform or mainly radial. However, in the presence of a rotating disk or a spinning black hole powering a jet, it could be a significant issue. To both evolve the magnetic field and avoid the severe time-step constraint, we opt for nested mesh with AMR instead of using a set of spherical annuli as in \citet{Cho2023ApJ...959L..22C, Cho2024ApJ...977..200C}. These key differences help us to handle a range of applications involving strong magnetic field or rotation (e.g., torus accretion; see \sect\ref{subsec:valid_torus}).

For the magnetized Bondi accretion onto a non-spinning black hole over galactic scales, we obtain an accretion rate and feedback power similar to those found by \citet{Cho2023ApJ...959L..22C} and \citet{Cho2024ApJ...977..200C}, providing strong validation of both methods for this problem.

\subsection{Future Applications}
The cyclic zoom method opens a window for various applications in the future to address increasingly realistic and complex problems. Of particular interest are investigations of Bondi accretion and torus accretion onto SMBHs coupled to galactic scales. In the presence of gas cooling and a large-scale gravitational potential from a galaxy, we can apply this method to multiphase accretion from galactic scales but using ab initio calculations instead of ad hoc models in current simulation~\citep[e.g.,][]{Fournier2024A&A...691A.239F, Grete2025arXiv250213213G}. More realistically, we could use this method to perform simulations using galaxies from cosmological simulations like the IllustrisTNG project~\citep{Pillepich2018MNRAS.473.4077P, Springel2018MNRAS.475..676S} instead of using current analytic or semi-analytic models for accretion and feedback ~\citep[e.g.,][]{Weinberger2017MNRAS.465.3291W, Weinberger2025arXiv250213241W}. In addition, the method can probably be generalized to GR-radiation MHD ~\citep{White2023ApJ...949..103W, Stone2024arXiv240916053S, Zhang2025arXiv250602289Z} and/or non-ideal GRMHD~\citep{Ripperda2019ApJS..244...10R}. In both cases, quantitative comparisons to fully resolved simulations to test the method will be necessary. More broadly, the method holds a lot of promise for being applied to many other problems where there are vast spatial and temporal scales to cover, e.g., convection in stars, collapsars, and fallback accretion.

\section{Summary}\label{sec:summary}
In this work, we have introduced the cyclic zoom method to resolve the dynamics of the accretion flow onto a black hole over a vast range of spatial and temporal scales in 3D GRMHD simulations. In this approach, we cyclically zoom out (derefine) and zoom in (refine) the simulation domain with a mask region in the center to preserve the small-scale physics and mediate an information flow from small to large scales and vice versa. The method can accelerate the GRMHD simulations by more than a factor of $\sim 10^5$.

We establish the validity of the method using a series of test problems including spherically symmetric Bondi accretion, monopole spin-down of a rotating black hole, magnetized turbulent Bondi accretion, torus accretion, and long-term secular evolution of a torus around both Schwarzschild and Kerr black holes. As applications, we simulate Bondi accretion and torus accretion of black holes from galactic scales onto both non-spinning and spinning black holes. For Bondi accretion, we find that the density scales with radius as $\rho\propto r^{-1}$ inside the Bondi radius. The accretion rate is suppressed relative to the Bondi rate by $\sim(10\,r_\mathrm{g}/r_\mathrm{B})^{1/2}$. We find energy feedback to the Bondi scale with a flux of $\sim 0.01 \dot{M} c^2$ for spin $a=0$, and $0.1 \dot{M} c^2$ for $a\approx0.9$. For the accretion of a torus, our model is able to capture the correct feedback efficiency and long-term secular evolution. We find that the accretion rate of a torus decreases with time as $\dot{M}\propto t^{-2}$ when the timescale is much longer than the viscous timescale.

By connecting physics at various scales, our approach aims to bridge the gap between galaxy scales and the event horizon of black holes. This method may lead to a more comprehensive subgrid model of black hole growth and feedback in large-scale galaxy formation cosmological simulations. The technique can furthermore be generalized to many other problems where there is a vast range of spatial and temporal scales to cover.

\section*{acknowledgments}
We thank Hyerin Cho, Ramesh Narayan, Alexander Tchekhovskoy, and Chang-Goo Kim for many useful conversations. We thank the anonymous referee for the helpful comments and suggestions. This work was supported by a grant from the Simons Foundation (888968, E.C. Ostriker, Princeton University PI) as part of the Learning the Universe Collaboration. J.S. acknowledges support from the Eric and Wendy Schmidt Fund for Strategic Innovation. E.Q. was supported in part by a Simons Investigator grant from the Simons Foundation and NSF AST grant 2107872.
We acknowledge the EuroHPC Joint Undertaking for awarding this project access to the EuroHPC supercomputer LUMI, hosted by CSC (Finland) and the LUMI consortium through a EuroHPC Regular Access call. 
The authors are pleased to acknowledge that the work reported on in this paper was substantially performed using the Princeton Research Computing resources at Princeton University, which is consortium of groups led by the Princeton Institute for Computational Science and Engineering (PICSciE) and Office of Information Technology's Research Computing.
This work used the Delta system at the National Center for Supercomputing Applications through allocation PHY230165 from the Advanced Cyberinfrastructure Coordination Ecosystem: Services \& Support (ACCESS) program, which is supported by National Science Foundation grants \#2138259, \#2138286, \#2138307, \#2137603, and \#2138296~\citep{Boerner10.1145}.

\vspace{5mm}
\software{\athenak{} \citep{Stone2024arXiv240916053S}}

\appendix

\section{Conservation of mass} \label{app:cons}

\begin{figure}[t!]
\includegraphics[width=\linewidth]{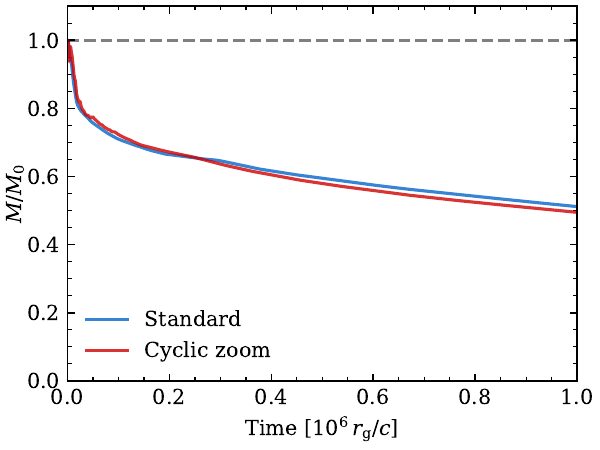}
\caption{History of the total mass in the simulation domain of the standard run and the cyclic zoom run of the long-term evolution of torus accretion with $a=0$. The relative difference is $\lesssim 1\%$.
\label{fig:mass_cons}}
\end{figure}

As we mentioned in \sect\ref{sec:method}, the cyclic zoom method is no longer conservative to machine precision, since the hydrodynamic variables are not evolved in the mask region (note, however, that magnetic flux is conserved, because we evolve magnetic fields inside the mask). In order to assess how significant the lack of conservation is,  \figu\ref{fig:mass_cons} shows the mass history for the long-term accretion of a rotating torus around a non-spinning black hole presented in \sect\ref{subsec:valid_secular}. At the end of the simulation ($t=10^6\,t_\mathrm{g}$), 50\% of the initial mass is left in the simulation domain, with 40\% accreted and 10\% escaping the outer boundary (see also \figu\ref{fig:secular}). The relative difference of mass remaining in the domain is $\lesssim 1\%$. Note that this is not an actual measure of the mass lost or gained in the cyclic zoom model, but rather just the difference between the evolution of the two models (which, as we argue in \sect\ref{subsec:diff}, should always be expected). Instead, the figure shows that the amount of mass lost or gained in the cyclic zoom model is negligible compared to the overall mass evolution of the torus due to dynamics. 

\bibliography{main}{}

\begin{thebibliography}{}
\expandafter\ifx\csname natexlab\endcsname\relax\def\natexlab#1{#1}\fi
\providecommand{\url}[1]{\href{#1}{#1}}
\providecommand{\dodoi}[1]{doi:~\href{http://doi.org/#1}{\nolinkurl{#1}}}
\providecommand{\doeprint}[1]{\href{http://ascl.net/#1}{\nolinkurl{http://ascl.net/#1}}}
\providecommand{\doarXiv}[1]{\href{https://arxiv.org/abs/#1}{\nolinkurl{https://arxiv.org/abs/#1}}}

\bibitem[{{Angl{\'e}s-Alc{\'a}zar} {et~al.}(2021){Angl{\'e}s-Alc{\'a}zar},
  {Quataert}, {Hopkins}, {Somerville}, {Hayward}, {Faucher-Gigu{\`e}re},
  {Bryan}, {Kere{\v{s}}}, {Hernquist}, \& {Stone}}]{Angles2021ApJ...917...53A}
{Angl{\'e}s-Alc{\'a}zar}, D., {Quataert}, E., {Hopkins}, P.~F., {et~al.} 2021,
  \apj, 917, 53, \dodoi{10.3847/1538-4357/ac09e8}

\bibitem[{{Blandford} \& {Znajek}(1977)}]{Blandford1977MNRAS.179..433B}
{Blandford}, R.~D., \& {Znajek}, R.~L. 1977, \mnras, 179, 433,
  \dodoi{10.1093/mnras/179.3.433}

\bibitem[{Boerner {et~al.}(2023)Boerner, Deems, Furlani, Knuth, \&
  Towns}]{Boerner10.1145}
Boerner, T.~J., Deems, S., Furlani, T.~R., Knuth, S.~L., \& Towns, J. 2023, in
  Practice and Experience in Advanced Research Computing, PEARC '23 (New York,
  NY, USA: Association for Computing Machinery), 173--176,
  \dodoi{10.1145/3569951.3597559}

\bibitem[{{Bondi}(1952)}]{Bondi1952MNRAS.112..195B}
{Bondi}, H. 1952, \mnras, 112, 195, \dodoi{10.1093/mnras/112.2.195}

\bibitem[{{Chael}(2024)}]{Chael2024MNRAS.532.3198C}
{Chael}, A. 2024, \mnras, 532, 3198, \dodoi{10.1093/mnras/stae1692}

\bibitem[{{Chael}(2025)}]{Chael2025MNRAS.537.2496C}
---. 2025, \mnras, 537, 2496, \dodoi{10.1093/mnras/staf200}

\bibitem[{{Cho} {et~al.}(2023){Cho}, {Prather}, {Narayan}, {Natarajan}, {Su},
  {Ricarte}, \& {Chatterjee}}]{Cho2023ApJ...959L..22C}
{Cho}, H., {Prather}, B.~S., {Narayan}, R., {et~al.} 2023, \apjl, 959, L22,
  \dodoi{10.3847/2041-8213/ad1048}

\bibitem[{{Cho} {et~al.}(2024){Cho}, {Prather}, {Su}, {Narayan}, \&
  {Natarajan}}]{Cho2024ApJ...977..200C}
{Cho}, H., {Prather}, B.~S., {Su}, K.-Y., {Narayan}, R., \& {Natarajan}, P.
  2024, \apj, 977, 200, \dodoi{10.3847/1538-4357/ad9561}

\bibitem[{{Event Horizon Telescope Collaboration} {et~al.}(2019){Event Horizon
  Telescope Collaboration}, {Akiyama}, {Alberdi}, {Alef}, {Asada}, {Azulay},
  {Baczko}, {Ball}, {Balokovi{\'c}}, \&
  {Barrett}}]{M87EHT_I_2019ApJ...875L...1E}
{Event Horizon Telescope Collaboration}, {Akiyama}, K., {Alberdi}, A., {et~al.}
  2019, \apjl, 875, L1, \dodoi{10.3847/2041-8213/ab0ec7}

\bibitem[{{Ferrarese} \& {Merritt}(2000)}]{Ferrarese2000ApJ...539L...9F}
{Ferrarese}, L., \& {Merritt}, D. 2000, \apjl, 539, L9, \dodoi{10.1086/312838}

\bibitem[{{Fishbone} \& {Moncrief}(1976)}]{Fishbone1976ApJ...207..962F}
{Fishbone}, L.~G., \& {Moncrief}, V. 1976, \apj, 207, 962,
  \dodoi{10.1086/154565}

\bibitem[{{Fournier} {et~al.}(2024){Fournier}, {Grete}, {Br{\"u}ggen},
  {Glines}, \& {O'Shea}}]{Fournier2024A&A...691A.239F}
{Fournier}, M., {Grete}, P., {Br{\"u}ggen}, M., {Glines}, F.~W., \& {O'Shea},
  B.~W. 2024, \aap, 691, A239, \dodoi{10.1051/0004-6361/202451031}

\bibitem[{{Galishnikova} {et~al.}(2025){Galishnikova}, {Philippov}, {Quataert},
  {Chatterjee}, \& {Liska}}]{Galishnikova2025ApJ...978..148G}
{Galishnikova}, A., {Philippov}, A., {Quataert}, E., {Chatterjee}, K., \&
  {Liska}, M. 2025, \apj, 978, 148, \dodoi{10.3847/1538-4357/ad9926}

\bibitem[{{Gammie} {et~al.}(2003){Gammie}, {McKinney}, \&
  {T{\'o}th}}]{Gammie2003ApJ...589..444G}
{Gammie}, C.~F., {McKinney}, J.~C., \& {T{\'o}th}, G. 2003, \apj, 589, 444,
  \dodoi{10.1086/374594}

\bibitem[{{Gaspari} {et~al.}(2020){Gaspari}, {Tombesi}, \&
  {Cappi}}]{Gaspari2020NatAs...4...10G}
{Gaspari}, M., {Tombesi}, F., \& {Cappi}, M. 2020, Nature Astronomy, 4, 10,
  \dodoi{10.1038/s41550-019-0970-1}

\bibitem[{{Gebhardt} {et~al.}(2000){Gebhardt}, {Bender}, {Bower}, {Dressler},
  {Faber}, {Filippenko}, {Green}, {Grillmair}, {Ho}, {Kormendy}, {Lauer},
  {Magorrian}, {Pinkney}, {Richstone}, \&
  {Tremaine}}]{Gebhardt2000ApJ...539L..13G}
{Gebhardt}, K., {Bender}, R., {Bower}, G., {et~al.} 2000, \apjl, 539, L13,
  \dodoi{10.1086/312840}

\bibitem[{{Grete} {et~al.}(2025){Grete}, {O'Shea}, {Glines}, {Prasad},
  {Wibking}, {Fournier}, {Br{\"u}ggen}, \& {Voit}}]{Grete2025arXiv250213213G}
{Grete}, P., {O'Shea}, B.~W., {Glines}, F.~W., {et~al.} 2025, arXiv e-prints,
  arXiv:2502.13213.
\newblock \doarXiv{2502.13213}

\bibitem[{{Guo} {et~al.}(2020){Guo}, {Inayoshi}, {Michiyama}, \&
  {Ho}}]{Guo2020ApJ...901...39G}
{Guo}, M., {Inayoshi}, K., {Michiyama}, T., \& {Ho}, L.~C. 2020, \apj, 901, 39,
  \dodoi{10.3847/1538-4357/abacc1}

\bibitem[{{Guo} {et~al.}(2023){Guo}, {Stone}, {Kim}, \&
  {Quataert}}]{Guo2023ApJ...946...26G}
{Guo}, M., {Stone}, J.~M., {Kim}, C.-G., \& {Quataert}, E. 2023, \apj, 946, 26,
  \dodoi{10.3847/1538-4357/acb81e}

\bibitem[{{Guo} {et~al.}(2024){Guo}, {Stone}, {Quataert}, \&
  {Kim}}]{Guo2024ApJ...973..141G}
{Guo}, M., {Stone}, J.~M., {Quataert}, E., \& {Kim}, C.-G. 2024, \apj, 973,
  141, \dodoi{10.3847/1538-4357/ad5fe7}

\bibitem[{{Hawley} {et~al.}(1984){Hawley}, {Smarr}, \&
  {Wilson}}]{Hawley1984ApJ...277..296H}
{Hawley}, J.~F., {Smarr}, L.~L., \& {Wilson}, J.~R. 1984, \apj, 277, 296,
  \dodoi{10.1086/161696}

\bibitem[{{Hopkins} \& {Quataert}(2010)}]{Hopkins&Quataert2010MNRAS.407.1529H}
{Hopkins}, P.~F., \& {Quataert}, E. 2010, \mnras, 407, 1529,
  \dodoi{10.1111/j.1365-2966.2010.17064.x}

\bibitem[{{Hopkins} \& {Quataert}(2011)}]{Hopkins&Quataert2011MNRAS.415.1027H}
---. 2011, \mnras, 415, 1027, \dodoi{10.1111/j.1365-2966.2011.18542.x}

\bibitem[{{Hopkins} {et~al.}(2024{\natexlab{a}}){Hopkins}, {Grudic}, {Su},
  {Wellons}, {Angles-Alcazar}, {Steinwandel}, {Guszejnov}, {Murray},
  {Faucher-Giguere}, {Quataert}, \& {Keres}}]{Hopkins2024OJAp....7E..18H}
{Hopkins}, P.~F., {Grudic}, M.~Y., {Su}, K.-Y., {et~al.} 2024{\natexlab{a}},
  The Open Journal of Astrophysics, 7, 18, \dodoi{10.21105/astro.2309.13115}

\bibitem[{{Hopkins} {et~al.}(2024{\natexlab{b}}){Hopkins}, {Squire}, {Su},
  {Steinwandel}, {Kremer}, {Shi}, {Grudic}, {Wellons}, {Faucher-Giguere},
  {Angles-Alcazar}, {Murray}, \& {Quataert}}]{Hopkins2024OJAp....7E..19H}
{Hopkins}, P.~F., {Squire}, J., {Su}, K.-Y., {et~al.} 2024{\natexlab{b}}, The
  Open Journal of Astrophysics, 7, 19, \dodoi{10.21105/astro.2310.04506}

\bibitem[{{Hopkins} {et~al.}(2025){Hopkins}, {Su}, {Murray}, {Steinwandel},
  {Kaaz}, {Ponnada}, {Bardati}, {Piotrowska}, {Wang}, {Shi}, {Angles-Alcazar},
  {Most}, {Kremer}, {Faucher-Giguere}, \&
  {Wellons}}]{Hopkins2025OJAp....8E..48H}
{Hopkins}, P.~F., {Su}, K.-Y., {Murray}, N., {et~al.} 2025, The Open Journal of
  Astrophysics, 8, 48, \dodoi{10.33232/001c.137296}

\bibitem[{{Kaaz} {et~al.}(2025){Kaaz}, {Liska}, {Tchekhovskoy}, {Hopkins}, \&
  {Jacquemin-Ide}}]{Kaaz2025ApJ...979..248K}
{Kaaz}, N., {Liska}, M., {Tchekhovskoy}, A., {Hopkins}, P.~F., \&
  {Jacquemin-Ide}, J. 2025, \apj, 979, 248, \dodoi{10.3847/1538-4357/ad9a86}

\bibitem[{{Kaaz} {et~al.}(2023){Kaaz}, {Murguia-Berthier}, {Chatterjee},
  {Liska}, \& {Tchekhovskoy}}]{Kaaz2023ApJ...950...31K}
{Kaaz}, N., {Murguia-Berthier}, A., {Chatterjee}, K., {Liska}, M. T.~P., \&
  {Tchekhovskoy}, A. 2023, \apj, 950, 31, \dodoi{10.3847/1538-4357/acc7a1}

\bibitem[{{Kormendy} \& {Ho}(2013)}]{Kormendy&Ho2013ARA&A..51..511K}
{Kormendy}, J., \& {Ho}, L.~C. 2013, \araa, 51, 511,
  \dodoi{10.1146/annurev-astro-082708-101811}

\bibitem[{{Koudmani} {et~al.}(2024){Koudmani}, {Somerville}, {Sijacki},
  {Bourne}, {Jiang}, \& {Profit}}]{Koudmani2024MNRAS.532...60K}
{Koudmani}, S., {Somerville}, R.~S., {Sijacki}, D., {et~al.} 2024, \mnras, 532,
  60, \dodoi{10.1093/mnras/stae1422}

\bibitem[{{Kozlowski} {et~al.}(1978){Kozlowski}, {Jaroszynski}, \&
  {Abramowicz}}]{Kozlowski1978A&A....63..209K}
{Kozlowski}, M., {Jaroszynski}, M., \& {Abramowicz}, M.~A. 1978, \aap, 63, 209

\bibitem[{{Lalakos} {et~al.}(2024){Lalakos}, {Tchekhovskoy}, {Bromberg},
  {Gottlieb}, {Jacquemin-Ide}, {Liska}, \&
  {Zhang}}]{Lalakos2024ApJ...964...79L}
{Lalakos}, A., {Tchekhovskoy}, A., {Bromberg}, O., {et~al.} 2024, \apj, 964,
  79, \dodoi{10.3847/1538-4357/ad0974}

\bibitem[{{Lalakos} {et~al.}(2022){Lalakos}, {Gottlieb}, {Kaaz}, {Chatterjee},
  {Liska}, {Christie}, {Tchekhovskoy}, {Zhuravleva}, \&
  {Nokhrina}}]{Lalakos2022ApJ...936L...5L}
{Lalakos}, A., {Gottlieb}, O., {Kaaz}, N., {et~al.} 2022, \apjl, 936, L5,
  \dodoi{10.3847/2041-8213/ac7bed}

\bibitem[{{Lemaster} \& {Stone}(2009)}]{Lemaster2009ApJ...691.1092L}
{Lemaster}, M.~N., \& {Stone}, J.~M. 2009, \apj, 691, 1092,
  \dodoi{10.1088/0004-637X/691/2/1092}

\bibitem[{{Li} \& {Bryan}(2014)}]{Li&Bryan2014ApJ...789..153L}
{Li}, Y., \& {Bryan}, G.~L. 2014, \apj, 789, 153,
  \dodoi{10.1088/0004-637X/789/2/153}

\bibitem[{{Magorrian} {et~al.}(1998){Magorrian}, {Tremaine}, {Richstone},
  {Bender}, {Bower}, {Dressler}, {Faber}, {Gebhardt}, {Green}, {Grillmair},
  {Kormendy}, \& {Lauer}}]{Magorrian1998AJ....115.2285M}
{Magorrian}, J., {Tremaine}, S., {Richstone}, D., {et~al.} 1998, \aj, 115,
  2285, \dodoi{10.1086/300353}

\bibitem[{{Metzger} {et~al.}(2008){Metzger}, {Piro}, \&
  {Quataert}}]{Metzger2008MNRAS.390..781M}
{Metzger}, B.~D., {Piro}, A.~L., \& {Quataert}, E. 2008, \mnras, 390, 781,
  \dodoi{10.1111/j.1365-2966.2008.13789.x}

\bibitem[{{Michel}(1972)}]{Michel1972Ap&SS..15..153M}
{Michel}, F.~C. 1972, \apss, 15, 153, \dodoi{10.1007/BF00649949}

\bibitem[{{Narayan} {et~al.}(2022){Narayan}, {Chael}, {Chatterjee}, {Ricarte},
  \& {Curd}}]{Narayan2022MNRAS.511.3795N}
{Narayan}, R., {Chael}, A., {Chatterjee}, K., {Ricarte}, A., \& {Curd}, B.
  2022, \mnras, 511, 3795, \dodoi{10.1093/mnras/stac285}

\bibitem[{{Narayan} {et~al.}(2012){Narayan}, {S{\"A} dowski}, {Penna}, \&
  {Kulkarni}}]{Narayan2012MNRAS.426.3241N}
{Narayan}, R., {S{\"A} dowski}, A., {Penna}, R.~F., \& {Kulkarni}, A.~K. 2012,
  \mnras, 426, 3241, \dodoi{10.1111/j.1365-2966.2012.22002.x}

\bibitem[{{Ni} {et~al.}(2024){Ni}, {Chen}, {Zhou}, {Park}, {Yang}, {DiMatteo},
  {Bird}, \& {Croft}}]{Ni2024arXiv240910666N}
{Ni}, Y., {Chen}, N., {Zhou}, Y., {et~al.} 2024, arXiv e-prints,
  arXiv:2409.10666, \dodoi{10.48550/arXiv.2409.10666}

\bibitem[{{Ni} {et~al.}(2022){Ni}, {Di Matteo}, {Bird}, {Croft}, {Feng},
  {Chen}, {Tremmel}, {DeGraf}, \& {Li}}]{Ni2022MNRAS.513..670N}
{Ni}, Y., {Di Matteo}, T., {Bird}, S., {et~al.} 2022, \mnras, 513, 670,
  \dodoi{10.1093/mnras/stac351}

\bibitem[{{Olivares} {et~al.}(2023){Olivares}, {Mo{\'s}cibrodzka}, \&
  {Porth}}]{Olivares2023A&A...678A.141O}
{Olivares}, H.~R., {Mo{\'s}cibrodzka}, M.~A., \& {Porth}, O. 2023, \aap, 678,
  A141, \dodoi{10.1051/0004-6361/202346010}

\bibitem[{{Pillepich} {et~al.}(2018){Pillepich}, {Springel}, {Nelson}, {Genel},
  {Naiman}, {Pakmor}, {Hernquist}, {Torrey}, {Vogelsberger}, {Weinberger}, \&
  {Marinacci}}]{Pillepich2018MNRAS.473.4077P}
{Pillepich}, A., {Springel}, V., {Nelson}, D., {et~al.} 2018, \mnras, 473,
  4077, \dodoi{10.1093/mnras/stx2656}

\bibitem[{{Porth} {et~al.}(2019){Porth}, {Chatterjee}, {Narayan}, {Gammie},
  {Mizuno}, {Anninos}, {Baker}, {Bugli}, {Chan}, {Davelaar}, {Del Zanna},
  {Etienne}, {Fragile}, {Kelly}, {Liska}, {Markoff}, {McKinney}, {Mishra},
  {Noble}, {Olivares}, {Prather}, {Rezzolla}, {Ryan}, {Stone}, {Tomei},
  {White}, {Younsi}, {Akiyama}, {Alberdi}, {Alef}, {Asada}, {Azulay}, {Baczko},
  {Ball}, {Balokovi{\'c}}, {Barrett}, {Bintley}, {Blackburn}, {Boland},
  {Bouman}, {Bower}, {Bremer}, {Brinkerink}, {Brissenden}, {Britzen},
  {Broderick}, {Broguiere}, {Bronzwaer}, {Byun}, {Carlstrom}, {Chael},
  {Chatterjee}, {Chen}, {Chen}, {Cho}, {Christian}, {Conway}, {Cordes},
  {Geoffrey}, {Crew}, {Cui}, {De Laurentis}, {Deane}, {Dempsey}, {Desvignes},
  {Doeleman}, {Eatough}, {Falcke}, {Fish}, {Fomalont}, {Fraga-Encinas},
  {Freeman}, {Friberg}, {Fromm}, {G{\'o}mez}, {Galison}, {Garc{\'\i}a},
  {Gentaz}, {Georgiev}, {Goddi}, {Gold}, {Gu}, {Gurwell}, {Hada}, {Hecht},
  {Hesper}, {Ho}, {Ho}, {Honma}, {Huang}, {Huang}, {Hughes}, {Ikeda}, {Inoue},
  {Issaoun}, {James}, {Jannuzi}, {Janssen}, {Jeter}, {Jiang}, {Johnson},
  {Jorstad}, {Jung}, {Karami}, {Karuppusamy}, {Kawashima}, {Keating},
  {Kettenis}, {Kim}, {Kim}, {Kim}, {Kino}, {Koay}, {Patrick}, {Koch}, {Koyama},
  {Kramer}, {Kramer}, {Krichbaum}, {Kuo}, {Lauer}, {Lee}, {Li}, {Li},
  {Lindqvist}, {Liu}, {Liuzzo}, {Lo}, {Lobanov}, {Loinard}, {Lonsdale}, {Lu},
  {MacDonald}, {Mao}, {Marrone}, {Marscher}, {Mart{\'\i}-Vidal}, {Matsushita},
  {Matthews}, {Medeiros}, {Menten}, {Mizuno}, {Moran}, {Moriyama},
  {Moscibrodzka}, {M{\"u}ller}, {Nagai}, {Nagar}, {Nakamura}, {Narayanan},
  {Natarajan}, {Neri}, {Ni}, {Noutsos}, {Okino}, {Oyama}, {{\"O}zel},
  {Palumbo}, {Patel}, {Pen}, {Pesce}, {Pi{\'e}tu}, {Plambeck}, {PopStefanija},
  {Preciado-L{\'o}pez}, {Psaltis}, {Pu}, {Ramakrishnan}, {Rao}, {Rawlings},
  {Raymond}, {Ripperda}, {Roelofs}, {Rogers}, {Ros}, {Rose}, {Roshanineshat},
  {Rottmann}, {Roy}, {Ruszczyk}, {Rygl}, {S{\'a}nchez},
  {S{\'a}nchez-Arguelles}, {Sasada}, {Savolainen}, {Schloerb}, {Schuster},
  {Shao}, {Shen}, {Small}, {Sohn}, {SooHoo}, {Tazaki}, {Tiede}, {Tilanus},
  {Titus}, {Toma}, {Torne}, {Trent}, {Trippe}, {Tsuda}, {van Bemmel}, {van
  Langevelde}, {van Rossum}, {Wagner}, {Wardle}, {Weintroub}, {Wex}, {Wharton},
  {Wielgus}, {Wong}, {Wu}, {Young}, {Young}, {Yuan}, {Yuan}, {Zensus}, {Zhao},
  {Zhao}, {Zhu}, \& {Event Horizon Telescope
  Collaboration}}]{Porth2019ApJS..243...26P}
{Porth}, O., {Chatterjee}, K., {Narayan}, R., {et~al.} 2019, \apjs, 243, 26,
  \dodoi{10.3847/1538-4365/ab29fd}

\bibitem[{{Ressler} {et~al.}(2018){Ressler}, {Quataert}, \&
  {Stone}}]{Ressler2018MNRAS.478.3544R}
{Ressler}, S.~M., {Quataert}, E., \& {Stone}, J.~M. 2018, \mnras, 478, 3544,
  \dodoi{10.1093/mnras/sty1146}

\bibitem[{{Ressler} {et~al.}(2020){Ressler}, {White}, {Quataert}, \&
  {Stone}}]{Ressler2020ApJ...896L...6R}
{Ressler}, S.~M., {White}, C.~J., {Quataert}, E., \& {Stone}, J.~M. 2020,
  \apjl, 896, L6, \dodoi{10.3847/2041-8213/ab9532}

\bibitem[{{Ripperda} {et~al.}(2019){Ripperda}, {Bacchini}, {Porth}, {Most},
  {Olivares}, {Nathanail}, {Rezzolla}, {Teunissen}, \&
  {Keppens}}]{Ripperda2019ApJS..244...10R}
{Ripperda}, B., {Bacchini}, F., {Porth}, O., {et~al.} 2019, \apjs, 244, 10,
  \dodoi{10.3847/1538-4365/ab3922}

\bibitem[{{Russell} {et~al.}(2015){Russell}, {Fabian}, {McNamara}, \&
  {Broderick}}]{Russell2015MNRAS.451..588R}
{Russell}, H.~R., {Fabian}, A.~C., {McNamara}, B.~R., \& {Broderick}, A.~E.
  2015, \mnras, 451, 588, \dodoi{10.1093/mnras/stv954}

\bibitem[{{Springel} {et~al.}(2018){Springel}, {Pakmor}, {Pillepich},
  {Weinberger}, {Nelson}, {Hernquist}, {Vogelsberger}, {Genel}, {Torrey},
  {Marinacci}, \& {Naiman}}]{Springel2018MNRAS.475..676S}
{Springel}, V., {Pakmor}, R., {Pillepich}, A., {et~al.} 2018, \mnras, 475, 676,
  \dodoi{10.1093/mnras/stx3304}

\bibitem[{{Stone} {et~al.}(2020){Stone}, {Tomida}, {White}, \&
  {Felker}}]{Stone2020ApJS..249....4S}
{Stone}, J.~M., {Tomida}, K., {White}, C.~J., \& {Felker}, K.~G. 2020, \apjs,
  249, 4, \dodoi{10.3847/1538-4365/ab929b}

\bibitem[{{Stone} {et~al.}(2024){Stone}, {Mullen}, {Fielding}, {Grete}, {Guo},
  {Kempski}, {Most}, {White}, \& {Wong}}]{Stone2024arXiv240916053S}
{Stone}, J.~M., {Mullen}, P.~D., {Fielding}, D., {et~al.} 2024, arXiv e-prints,
  arXiv:2409.16053, \dodoi{10.48550/arXiv.2409.16053}

\bibitem[{{Su} {et~al.}(2023){Su}, {Bryan}, {Haiman}, {Somerville}, {Hayward},
  \& {Faucher-Gigu{\`e}re}}]{Su2023MNRAS.520.4258S}
{Su}, K.-Y., {Bryan}, G.~L., {Haiman}, Z., {et~al.} 2023, \mnras, 520, 4258,
  \dodoi{10.1093/mnras/stad252}

\bibitem[{{Tchekhovskoy} {et~al.}(2011){Tchekhovskoy}, {Narayan}, \&
  {McKinney}}]{Tchekhovskoy2011MNRAS.418L..79T}
{Tchekhovskoy}, A., {Narayan}, R., \& {McKinney}, J.~C. 2011, \mnras, 418, L79,
  \dodoi{10.1111/j.1745-3933.2011.01147.x}

\bibitem[{{Tremmel} {et~al.}(2017){Tremmel}, {Karcher}, {Governato},
  {Volonteri}, {Quinn}, {Pontzen}, {Anderson}, \&
  {Bellovary}}]{Tremmel2017MNRAS.470.1121T}
{Tremmel}, M., {Karcher}, M., {Governato}, F., {et~al.} 2017, \mnras, 470,
  1121, \dodoi{10.1093/mnras/stx1160}

\bibitem[{{Weinberger} {et~al.}(2025){Weinberger}, {Bhowmick}, {Blecha},
  {Bryan}, {Buchner}, {Hernquist}, {Hlavacek-Larrondo}, \&
  {Springel}}]{Weinberger2025arXiv250213241W}
{Weinberger}, R., {Bhowmick}, A., {Blecha}, L., {et~al.} 2025, arXiv e-prints,
  arXiv:2502.13241.
\newblock \doarXiv{2502.13241}

\bibitem[{{Weinberger} {et~al.}(2017){Weinberger}, {Springel}, {Hernquist},
  {Pillepich}, {Marinacci}, {Pakmor}, {Nelson}, {Genel}, {Vogelsberger},
  {Naiman}, \& {Torrey}}]{Weinberger2017MNRAS.465.3291W}
{Weinberger}, R., {Springel}, V., {Hernquist}, L., {et~al.} 2017, \mnras, 465,
  3291, \dodoi{10.1093/mnras/stw2944}

\bibitem[{{White} {et~al.}(2023){White}, {Mullen}, {Jiang}, {Davis}, {Stone},
  {Morozova}, \& {Zhang}}]{White2023ApJ...949..103W}
{White}, C.~J., {Mullen}, P.~D., {Jiang}, Y.-F., {et~al.} 2023, \apj, 949, 103,
  \dodoi{10.3847/1538-4357/acc8cf}

\bibitem[{{White} {et~al.}(2020){White}, {Quataert}, \&
  {Gammie}}]{White2020ApJ...891...63W}
{White}, C.~J., {Quataert}, E., \& {Gammie}, C.~F. 2020, \apj, 891, 63,
  \dodoi{10.3847/1538-4357/ab718e}

\bibitem[{{White} {et~al.}(2016){White}, {Stone}, \&
  {Gammie}}]{White2016ApJS..225...22W}
{White}, C.~J., {Stone}, J.~M., \& {Gammie}, C.~F. 2016, \apjs, 225, 22,
  \dodoi{10.3847/0067-0049/225/2/22}

\bibitem[{{Xu}(2023)}]{Xu2023ApJ...954..180X}
{Xu}, W. 2023, \apj, 954, 180, \dodoi{10.3847/1538-4357/ace892}

\bibitem[{{Xu} \& {Stone}(2019)}]{Xu2019MNRAS.488.5162X}
{Xu}, W., \& {Stone}, J.~M. 2019, \mnras, 488, 5162,
  \dodoi{10.1093/mnras/stz2002}

\bibitem[{{Yang} {et~al.}(2021){Yang}, {Yuan}, {Yuan}, \&
  {White}}]{Yang2021ApJ...914..131Y}
{Yang}, H., {Yuan}, F., {Yuan}, Y.-F., \& {White}, C.~J. 2021, \apj, 914, 131,
  \dodoi{10.3847/1538-4357/abfe63}

\bibitem[{{Yuan} \& {Narayan}(2014)}]{Yuan2014ARA&A..52..529Y}
{Yuan}, F., \& {Narayan}, R. 2014, \araa, 52, 529,
  \dodoi{10.1146/annurev-astro-082812-141003}

\bibitem[{{Zhang} {et~al.}(2025){Zhang}, {Stone}, {Mullen}, {Davis}, {Jiang},
  \& {White}}]{Zhang2025arXiv250602289Z}
{Zhang}, L., {Stone}, J.~M., {Mullen}, P.~D., {et~al.} 2025, arXiv e-prints,
  arXiv:2506.02289, \dodoi{10.48550/arXiv.2506.02289}

\end{thebibliography}
\bibliographystyle{aasjournal}


\end{CJK*}
\end{document}